\documentclass[aps,twocolumn,superscriptaddress,showpacs,preprintnumbers,amsmath,amssymb]{revtex4}
\usepackage{graphicx,color,dcolumn,booktabs,bm}
\usepackage{longtable,lscape}
\usepackage{amsmath}
\usepackage{indentfirst}
\usepackage{epsfig}
\usepackage{feynmf}   
\usepackage{epstopdf}   
\usepackage{slashed}  
\usepackage{cases}
\usepackage{color}
\usepackage{multirow}
\usepackage{graphicx,color,dcolumn,booktabs,bm}
\usepackage{verbatim}
\usepackage[colorlinks,linkcolor=red,anchorcolor=green,citecolor=blue]{hyperref}
\usepackage{mathrsfs}
\usepackage{rotating}
\usepackage{threeparttable}
\usepackage{lineno}

\newcommand{\BR}{{\cal B}}

\newcommand{\jpsi}{J/\psi}

\newcommand{\EE}{e^+e^-}
\newcommand{\MM}{\mu^+\mu^-}
\newcommand{\LL}{\ell^+\ell^-}

\newcommand{\beq}{\begin{equation}}
\newcommand{\eeq}{\end{equation}}
\newcommand{\bitm}{\begin{itemize}}
\newcommand{\eitm}{\end{itemize}}

\begin{document}


\title{\quad\\[2.0cm]\boldmath Search for
$\Upsilon(1S,2S) \to Z^{+}_{c}Z^{(\prime) -}_{c}$ and $e^{+}e^{-} \to
Z^{+}_{c}Z^{(\prime) -}_{c}$ at $\sqrt{s}$ = 10.52, 10.58, and 10.867 GeV}


\noaffiliation
\affiliation{University of the Basque Country UPV/EHU, 48080 Bilbao}
\affiliation{Beihang University, Beijing 100191}
\affiliation{Brookhaven National Laboratory, Upton, New York 11973}
\affiliation{Budker Institute of Nuclear Physics SB RAS, Novosibirsk 630090}
\affiliation{Faculty of Mathematics and Physics, Charles University, 121 16 Prague}
\affiliation{University of Cincinnati, Cincinnati, Ohio 45221}
\affiliation{Deutsches Elektronen--Synchrotron, 22607 Hamburg}
\affiliation{Duke University, Durham, North Carolina 27708}
\affiliation{University of Florida, Gainesville, Florida 32611}
\affiliation{Key Laboratory of Nuclear Physics and Ion-beam Application (MOE) and Institute of Modern Physics, Fudan University, Shanghai 200443}
\affiliation{Justus-Liebig-Universit\"at Gie\ss{}en, 35392 Gie\ss{}en}
\affiliation{Gifu University, Gifu 501-1193}
\affiliation{II. Physikalisches Institut, Georg-August-Universit\"at G\"ottingen, 37073 G\"ottingen}
\affiliation{SOKENDAI (The Graduate University for Advanced Studies), Hayama 240-0193}
\affiliation{Gyeongsang National University, Chinju 660-701}
\affiliation{Hanyang University, Seoul 133-791}
\affiliation{University of Hawaii, Honolulu, Hawaii 96822}
\affiliation{High Energy Accelerator Research Organization (KEK), Tsukuba 305-0801}
\affiliation{J-PARC Branch, KEK Theory Center, High Energy Accelerator Research Organization (KEK), Tsukuba 305-0801}
\affiliation{IKERBASQUE, Basque Foundation for Science, 48013 Bilbao}
\affiliation{Indian Institute of Technology Bhubaneswar, Satya Nagar 751007}
\affiliation{Indian Institute of Technology Guwahati, Assam 781039}
\affiliation{Indian Institute of Technology Hyderabad, Telangana 502285}
\affiliation{Indian Institute of Technology Madras, Chennai 600036}
\affiliation{Indiana University, Bloomington, Indiana 47408}
\affiliation{Institute of High Energy Physics, Chinese Academy of Sciences, Beijing 100049}
\affiliation{Institute of High Energy Physics, Vienna 1050}
\affiliation{Institute for High Energy Physics, Protvino 142281}
\affiliation{INFN - Sezione di Napoli, 80126 Napoli}
\affiliation{INFN - Sezione di Torino, 10125 Torino}
\affiliation{Advanced Science Research Center, Japan Atomic Energy Agency, Naka 319-1195}
\affiliation{J. Stefan Institute, 1000 Ljubljana}
\affiliation{Kanagawa University, Yokohama 221-8686}
\affiliation{Institut f\"ur Experimentelle Teilchenphysik, Karlsruher Institut f\"ur Technologie, 76131 Karlsruhe}
\affiliation{Kennesaw State University, Kennesaw, Georgia 30144}
\affiliation{King Abdulaziz City for Science and Technology, Riyadh 11442}
\affiliation{Department of Physics, Faculty of Science, King Abdulaziz University, Jeddah 21589}
\affiliation{Korea Institute of Science and Technology Information, Daejeon 305-806}
\affiliation{Korea University, Seoul 136-713}
\affiliation{Kyoto University, Kyoto 606-8502}
\affiliation{Kyungpook National University, Daegu 702-701}
\affiliation{LAL, Univ. Paris-Sud, CNRS/IN2P3, Universit\'{e} Paris-Saclay, Orsay}
\affiliation{\'Ecole Polytechnique F\'ed\'erale de Lausanne (EPFL), Lausanne 1015}
\affiliation{P.N. Lebedev Physical Institute of the Russian Academy of Sciences, Moscow 119991}
\affiliation{Faculty of Mathematics and Physics, University of Ljubljana, 1000 Ljubljana}
\affiliation{Ludwig Maximilians University, 80539 Munich}
\affiliation{Luther College, Decorah, Iowa 52101}
\affiliation{University of Malaya, 50603 Kuala Lumpur}
\affiliation{University of Maribor, 2000 Maribor}
\affiliation{Max-Planck-Institut f\"ur Physik, 80805 M\"unchen}
\affiliation{School of Physics, University of Melbourne, Victoria 3010}
\affiliation{University of Mississippi, University, Mississippi 38677}
\affiliation{University of Miyazaki, Miyazaki 889-2192}
\affiliation{Moscow Physical Engineering Institute, Moscow 115409}
\affiliation{Moscow Institute of Physics and Technology, Moscow Region 141700}
\affiliation{Graduate School of Science, Nagoya University, Nagoya 464-8602}
\affiliation{Kobayashi-Maskawa Institute, Nagoya University, Nagoya 464-8602}
\affiliation{Universit\`{a} di Napoli Federico II, 80055 Napoli}
\affiliation{Nara Women's University, Nara 630-8506}
\affiliation{National Central University, Chung-li 32054}
\affiliation{National United University, Miao Li 36003}
\affiliation{Department of Physics, National Taiwan University, Taipei 10617}
\affiliation{H. Niewodniczanski Institute of Nuclear Physics, Krakow 31-342}
\affiliation{Nippon Dental University, Niigata 951-8580}
\affiliation{Niigata University, Niigata 950-2181}
\affiliation{University of Nova Gorica, 5000 Nova Gorica}
\affiliation{Novosibirsk State University, Novosibirsk 630090}
\affiliation{Osaka City University, Osaka 558-8585}
\affiliation{Pacific Northwest National Laboratory, Richland, Washington 99352}
\affiliation{Panjab University, Chandigarh 160014}
\affiliation{Peking University, Beijing 100871}
\affiliation{University of Pittsburgh, Pittsburgh, Pennsylvania 15260}
\affiliation{Theoretical Research Division, Nishina Center, RIKEN, Saitama 351-0198}
\affiliation{University of Science and Technology of China, Hefei 230026}
\affiliation{Showa Pharmaceutical University, Tokyo 194-8543}
\affiliation{Soongsil University, Seoul 156-743}
\affiliation{Stefan Meyer Institute for Subatomic Physics, Vienna 1090}
\affiliation{Sungkyunkwan University, Suwon 440-746}
\affiliation{Department of Physics, Faculty of Science, University of Tabuk, Tabuk 71451}
\affiliation{Tata Institute of Fundamental Research, Mumbai 400005}
\affiliation{Excellence Cluster Universe, Technische Universit\"at M\"unchen, 85748 Garching}
\affiliation{Department of Physics, Technische Universit\"at M\"unchen, 85748 Garching}
\affiliation{Toho University, Funabashi 274-8510}
\affiliation{Department of Physics, Tohoku University, Sendai 980-8578}
\affiliation{Earthquake Research Institute, University of Tokyo, Tokyo 113-0032}
\affiliation{Department of Physics, University of Tokyo, Tokyo 113-0033}
\affiliation{Tokyo Institute of Technology, Tokyo 152-8550}
\affiliation{Tokyo Metropolitan University, Tokyo 192-0397}
\affiliation{Virginia Polytechnic Institute and State University, Blacksburg, Virginia 24061}
\affiliation{Wayne State University, Detroit, Michigan 48202}
\affiliation{Yamagata University, Yamagata 990-8560}
\affiliation{Yonsei University, Seoul 120-749}
  \author{S.~Jia}\affiliation{Beihang University, Beijing 100191} 
  \author{C.~P.~Shen}\affiliation{Beihang University, Beijing 100191} 
  \author{C.~Z.~Yuan}\affiliation{Institute of High Energy Physics, Chinese Academy of Sciences, Beijing 100049} 
  \author{I.~Adachi}\affiliation{High Energy Accelerator Research Organization (KEK), Tsukuba 305-0801}\affiliation{SOKENDAI (The Graduate University for Advanced Studies), Hayama 240-0193} 
  \author{H.~Aihara}\affiliation{Department of Physics, University of Tokyo, Tokyo 113-0033} 
  \author{S.~Al~Said}\affiliation{Department of Physics, Faculty of Science, University of Tabuk, Tabuk 71451}\affiliation{Department of Physics, Faculty of Science, King Abdulaziz University, Jeddah 21589} 
  \author{D.~M.~Asner}\affiliation{Brookhaven National Laboratory, Upton, New York 11973} 
  \author{V.~Aulchenko}\affiliation{Budker Institute of Nuclear Physics SB RAS, Novosibirsk 630090}\affiliation{Novosibirsk State University, Novosibirsk 630090} 
  \author{T.~Aushev}\affiliation{Moscow Institute of Physics and Technology, Moscow Region 141700} 
  \author{R.~Ayad}\affiliation{Department of Physics, Faculty of Science, University of Tabuk, Tabuk 71451} 
% ...
\collaboration{The Belle Collaboration}


\begin{abstract}

The first search for double charged charmoniumlike state  production in $\Upsilon(1S)$
and $\Upsilon(2S)$ decays and in $e^{+}e^{-}$ annihilation at
$\sqrt{s}$ = 10.52, 10.58, and 10.867 GeV is conducted using
data collected with the Belle detector at the KEKB asymmetric
energy electron-positron collider. No significant signals are
observed in any of the studied modes, and the 90\% credibility level
upper limits on their product branching fractions in $\Upsilon(1S)$ and
$\Upsilon(2S)$ decays (${\cal B}(\Upsilon(1S,2S)\to
Z^{+}_{c}Z^{(\prime) -}_{c})\times{\cal B}(Z^{+}_{c}\to\pi^{+}+c\bar c)$ ($c\bar c=J/\psi$,
$\chi_{c1}(1P)$, $\psi(2S)$)) and the product of Born cross section
and branching fraction for $e^{+}e^{-} \to Z^{+}_{c}Z^{(\prime) -}_{c}$
($\sigma(e^+e^- \to Z^{+}_{c}Z^{(\prime) -}_{c})\times {\cal B}(Z^{+}_{c} \to
\pi^+ +c\bar c)$) at $\sqrt{s}$ = 10.52, 10.58, and 10.867 GeV are
determined.
Here, $Z_{c}$ refers to the $Z_{c}(3900)$ and
$Z_{c}(4200)$ observed in the $\pi\jpsi$ final state, the $Z_{c1}(4050)$ and
$Z_{c2}(4250)$ in the $\pi\chi_{c1}(1P)$ final state, and the $Z_{c}(4050)$ and
$Z_{c}(4430)$ in the $\pi\psi(2S)$ final state.

\end{abstract}

\pacs{14.40.Rt, 13.25.Gv, 14.40.Pq}

\maketitle

\section{\boldmath Introduction}

In the past decade, many experiments, at both lepton and
hadron colliders, reported evidences for a large number of new
particles having exotic properties that are difficult to accommodate within the conventional quark model, especially those that couple¡± to heavy
quarkonium and have non-zero charge~\cite{zhusl}. Since the
discovery of the $Z^{+}_{c}(4430)$($\to\pi^{+}\psi(2S)$)~\cite{100.142001}, more charged
charmoniumlike resonances have been observed, including
$Z^{+}_{c1}(4050)$ and $Z^{+}_{c2}(4250)$($\to\pi^{+}\chi_{c1}(1P)$)~\cite{D78.072004},
$Z^{+}_{c}(3900)$~\cite{110.252001,110.252002} and
$Z^{+}_{c}(4200)$($\to\pi^{+}\jpsi$)~\cite{D90.112009}, and
$Z^{+}_{c}(4050)$($\to\pi^{+}\psi(2S)$)~\cite{D91.112007}.
The preferred assignment of the quantum numbers for
$Z_c^+(4430)$, $Z_{c}^+(3900)$, and $Z_{c}^+(4200)$ are
$J^{P}=1^{+}$~\cite{112.222002,119.072001,D90.112009}. Although
these charged charmoniumlike candidates are widely interpreted
unconventional $c\bar c$ states, it remains a significant
challenge to determine their internal dynamics~\cite{1708.04012}.

Considerable efforts in theory have been devoted to interpret
these states as tetraquarks, molecules, hybrids, or
hadrocharmonia~\cite{C71.1534,C74.2981,A29.1430046}. To
distinguish among these explanations, experimental input is needed, such as that from the study of double
$Z^{\pm}_{c}$ production in $\EE$
annihilation~\cite{B764,D91.114025}.
For $e^{+}e^{-} \to Z^{+}_{c}Z^{-}_{c}$, the dependence on $s$ (the $e^+ e^-$ center-of-mass (CM) energy squared) of the electromagnetic form factor, $F_{Z_{c}^{+}Z_{c}^{-}}$ is $1/s^3$ for a
$Z_{c}$ state with tetraquark structure or $1/s$ for a $Z_c$ system of
two tightly bound diquarks~\cite{B764,D91.114025}.

In this paper, we report the search for $Z_{c}$ pair
production in $\Upsilon(1S)$ and $\Upsilon(2S)$ decays as well as
in $\EE$ annihilation at $\sqrt{s} =$ 10.52, 10.58, and 10.867~GeV.
We measure the production rates of $\Upsilon(1S)$ and
$\Upsilon(2S)$ decay into a pair of $Z_{c}$ states and the Born
cross sections of $e^{+}e^{-} \to Z^{+}_{c}Z^{(\prime) -}_{c}$ at
$\sqrt{s} =$ 10.52,~10.58, and 10.867~GeV. Here, $Z^{+}_{c}$
denotes $Z^{+}_{c}(3900)$, $Z^{+}_{c}(4200)$, $Z^{+}_{c1}(4050)$,
$Z^{+}_{c2}(4250)$, $Z^{+}_{c}(4050)$, or $Z^{+}_{c}(4430)$. In
these searches, the decay modes considered are
$Z^{+}_{c}(3900)/Z^{+}_{c}(4200) \to \pi^{+} \jpsi$,
$Z^{+}_{c1}(4050)/Z^{+}_{c2}(4250)\to\pi^{+} \chi_{c1}(1P)$ and
$Z^{+}_{c}(4050)/Z^{+}_{c}(4430) \to \pi^{+}
\psi(2S)$~\cite{charge}.
Although it is possible to search for combinations of any two of the enumerated $Z_c$ states, only
both Zc states decay into the same final states are currently studied in this study;
the information of the other final-state combinations can be estimated
from the results reported here.

\section{\boldmath The data sample and the belle detector}

This analysis utilizes ($5.74\pm0.09$)~fb$^{-1}$ data collected at the
$\Upsilon(1S)$ peak (($102\pm3$) million $\Upsilon(1S)$ events),
($24.91\pm0.35$)~fb$^{-1}$ data collected at the $\Upsilon(2S)$ peak (($158\pm4$)
million $\Upsilon(2S)$ events), ($89.5\pm1.3$)~fb$^{-1}$ data collected at
$\sqrt{s} = 10.52$~GeV, ($711.0\pm10.0$)~fb$^{-1}$ data collected at $\sqrt{s}
=10.58$~GeV ($\Upsilon$(4S) peak), and ($121.4\pm1.7$)~fb$^{-1}$ data
collected at $\sqrt{s} =10.867$~GeV ($\Upsilon$(5S) peak). All the
data are collected with the Belle detector~\cite{Belle} operating
at the KEKB asymmetric-energy $\EE$ collider~\cite{KEKB}.
The Belle detector is a large solid-angle magnetic spectrometer that
consists of a silicon vertex detector, a 50-layer central drift
chamber (CDC), an array of aerogel threshold Cherenkov counters (ACC), a
barrel-like arrangement of time-of-flight scintillation counters (TOF),
and an electromagnetic calorimeter comprised of CsI(TI) crystals (ECL)
located inside a superconducting solenoid coil that provides a
$1.5~\hbox{T}$ magnetic field. An iron flux-return yoke
instrumented with resistive plate chambers located outside the
coil is used to detect $K^{0}_{L}$ mesons and to identify muons. A
detailed description of the Belle detector can be found in
Refs.~\cite{Belle, PTEP201204D001}.

We generate large signal Monte Carlo (MC) samples with the {\sc
evtgen} event generator~\cite{EVTGEN} to understand the signal event topology and to estimate
the signal selection efficiency. The angular
distribution for $\Upsilon(1S,2S) \to Z^{+}_{c}Z^{(\prime) -}_{c}$ and
$e^{+}e^{-} \to Z^{+}_{c}Z^{(\prime) -}_{c}$ at $\sqrt{s} =$ 10.52,
10.58, and 10.867~GeV is simulated assuming an $e^{+}e^{-} \to A \bar
A$ decay mode ($A$ denoting an axial-vector state), \textit{i.e.,} $dN/d{\rm
cos} \theta_{Z_{c}} \propto 1-{\rm
cos}^2\theta_{Z_{c}}$~\cite{DPNU28}, where $\theta_{Z_{c}}$ is the
polar angle of the $Z_{c}$ in the $\EE$ CM system. Initial state
radiation (ISR) is taken into account by assuming that the
cross sections follow a $1/s^2$ dependence in $e^+e^-\to Z^{+}_{c}Z^{(\prime) -}_{c}$ reactions. For the $Z_c$ pair in one MC event,  one $Z_c$ decays into $\pi^{\pm} J/\psi$, $\pi^{\pm}\chi_{c1}(1P)$, or
$\pi^{\pm}\psi(2S)$ using the phase space model, while the other is simulated
with the inclusive decays using {\sc pythia}~\cite{PYTHIA}. The
masses and widths of the $Z^{\pm}_{c}$ states are set according to the
latest world-average values~\cite{C38.090001}, as summarized in Table~\ref{PDG}.

\begin{table}[htbp]
\caption{The world-average (nominal) values of $Z^{\pm}_{c}$ masses (in MeV/$c^2$) and widths (in MeV)~\cite{C38.090001}.}\label{PDG}
\begin{tabular}{c | c | c | c }
\hline
$Z_{c}$ states & $Z_{c}$ labels in Ref.~\cite{C38.090001} & Mass & Width\\
\hline
$Z^+_{c}(3900)$ & $X^+(3900)$ & $3886.6\pm2.4$ & $28.1\pm2.6$ \\
$Z^+_{c}(4200)$ & $X^+(4200)$ & $4196^{+35}_{-32}$ & $370^{+100}_{-150}$ \\
$Z^+_{c1}(4050)$ & $X^+(4050)$ & $4051^{+24}_{-40}$ & $82^{+50}_{-28}$ \\
$Z^+_{c2}(4250)$ & $X^+(4250)$ & $4248^{+190}_{-50}$ & $177^{+320}_{-70}$ \\
$Z^+_{c}(4050)$ & $X^+(4055)$ & $4054\pm3.2$ & $45\pm13$ \\
$Z^+_{c}(4430)$ & $X^+(4430)$ & $4478^{+15}_{-18}$ & $181\pm31$ \\
\hline
\hline
\end{tabular}
\end{table}

\section{\boldmath Common Event selection criteria}

For well reconstructed charged tracks, the impact parameters
perpendicular to and along the beam direction with respect to the
nominal interaction point are required to be less than 0.5 cm and
4 cm, respectively, and the transverse momentum in the laboratory
frame is required to be larger than 0.1~GeV/$c$. We require the
number of well reconstructed charged tracks to be greater than
four to suppress the significant background from quantum
electrodynamics processes. For charged tracks, information from
different detector subsystems including specific ionization in the
CDC, time measurements in the TOF and the response of the ACC, is
combined to form the likelihood ${\mathcal L}_i$ for particle
species $i$, where $i=\pi$,~$K$, or $p$~\cite{A494.402}. Charged
tracks with
$R_{K}=\mathcal{L}_{K}/(\mathcal{L}_K+\mathcal{L}_\pi)<0.4$ are
considered to be pions. With this condition, the pion
identification efficiency is $97\%$ and the kaon to pion misidentification
rate is about $4\%$. A similar likelihood ratio is defined as
$R_e=\mathcal{L}_e/(\mathcal{L}_e+\mathcal{L}_{{\rm
non}-e})$~\cite{Hanagaki485490} for electron identification and
$R_{\mu}=\mathcal{L}_{\mu}/(\mathcal{L}_{\mu}+\mathcal{L}_{K}
+\mathcal{L}_{\pi})$~\cite{Abashian49169} for muon identification.
An ECL cluster is treated as a photon candidate if it is isolated from
the projected path of charged tracks in the CDC, and its energy is greater
than 50~MeV.

For the lepton pair $\LL$ ($\ell=e$ or $\mu$) used to reconstruct
the $\jpsi$, both tracks must have $R_e>0.95$ in the
$\EE$ mode, while one track must have $R_\mu>0.95$ and the other
$R_\mu>0.05$ in the $\MM$ mode. The lepton pair identification
efficiencies for $e^+e^-$ and $\mu^+\mu^-$ are $96\%$ and $93\%$,
respectively. We apply a lepton veto to the bachelor pion candidate by requiring $R_\mu<0.95$ and
$R_e<0.95$. To reduce the effect of bremsstrahlung and final-state
radiation, photons detected in the ECL within a $50~\mathrm{mrad}$
cone of the original electron or positron direction are absorbed into the $e^+/e^-$ four-momentum.

\section{\boldmath Search for $\Upsilon(1S,2S) \to Z^{+}_{c}Z^{(\prime) -}_{c}$
and $e^{+}e^{-} \to Z^{+}_{c}Z^{(\prime) -}_{c}$ at $\sqrt{s} =$ 10.52,
10.58, and 10.867~GeV with $Z^{+}_{c} \to \pi^{+} \jpsi$}

We search for the production of double-$Z_{c}(3900)$ states,
double-$Z_{c}(4200)$ states, and $Z_{c}(3900)$-plus-$Z_{c}(4200)$
states in $\Upsilon(1S,2S)$ decays and $e^{+}e^{-}$ annihilation
at $\sqrt{s} =$ 10.52, 10.58, and 10.867~GeV, where one $Z_{c}$
decays into $\pi^{+}$ and $\jpsi$ ($\to \ell^+ \ell^-$), and the other
decays inclusively.

After applying the aforementioned common event selections, the invariant mass
distribution of the $\jpsi$ candidates from the five data samples are shown in
Figs.~\ref{mlldata}(a)--(e). Clear $\jpsi$ signals are
observed. The $\jpsi$ signal region is defined as
$|M_{\ell^+\ell^-}-m_{\jpsi}|<0.03$~GeV/$c^2$ ($>98\%$ signal events are reserved according to the MC simulation), where $m_{\jpsi}$ is the nominal
mass~\cite{C38.090001}, while the $\jpsi$ mass sidebands are
$2.97<M_{\ell^+\ell^-}<3.03$~GeV/$c^2$ or
$3.17<M_{\ell^+\ell^-}<3.23$~GeV/$c^2$ (twice the width of the signal region). In order to
improve the $\jpsi$ momentum resolution, a mass-constrained-fit is
applied to the $\jpsi$ candidates in the signal region.

\begin{figure*}[htbp]
\includegraphics[width=3.5cm,angle=-90]{fig1a.epsi}
\includegraphics[width=3.5cm,angle=-90]{fig1b.epsi}
\vspace{0.3cm}
\includegraphics[width=3.5cm,angle=-90]{fig1c.epsi}
\includegraphics[width=3.5cm,angle=-90]{fig1d.epsi}
\includegraphics[width=3.5cm,angle=-90]{fig1e.epsi}
\caption{The invariant mass distributions of the $\jpsi$
candidates  from (a) $\Upsilon(1S)$, (b) $\Upsilon(2S)$, (c)
$\sqrt{s} = 10.52$~GeV, (d) $\sqrt{s} = 10.58$~GeV, and (e)
$\sqrt{s} = 10.867$~GeV data samples. The solid red arrows show the
$\jpsi$ signal region, and the dashed black arrows show the $\jpsi$ mass
sideband regions.}\label{mlldata}
\end{figure*}

\begin{figure*}[htbp]
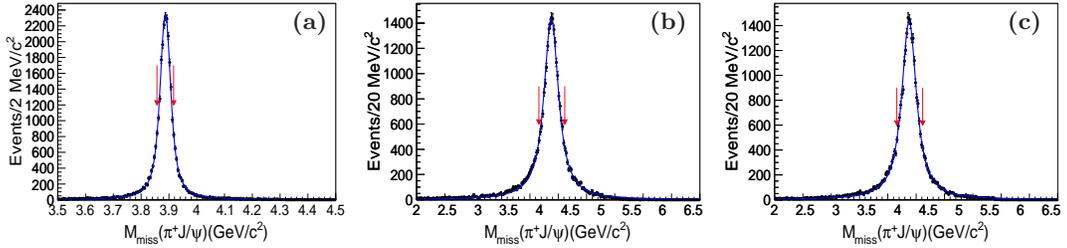

\includegraphics[height=4.5cm,width=3.2cm,angle=-90]{fig2a.epsi}
\put(-20, -10){\bf (a)}
\hspace{0.15cm}
\includegraphics[height=4.5cm,width=3.2cm,angle=-90]{fig2b.epsi}
\put(-20, -10){\bf (b)}
\hspace{0.15cm}
\includegraphics[height=4.5cm,width=3.2cm,angle=-90]{fig2c.epsi}
\put(-20, -10){\bf (c)}
\caption{The recoil mass spectra of the $Z^{+}_{c}$ ($\to \pi^+ J/\psi$) states from MC simulated (a) $\Upsilon(1S)\to Z^{+}_{c}(3900)Z^{-}_{c}(3900)$, (b) $\Upsilon(1S)\to Z^{+}_{c}(4200)Z^{-}_{c}(4200)$, and (c) $\Upsilon(1S)\to Z^{+}_{c}(3900)Z^{-}_{c}(4200)+c.c.$.}\label{MM1}
\end{figure*}

For the events with the lepton-pair mass within the $\jpsi$ signal region,
Figs.~\ref{MM1}(a)--(c) show the recoil mass spectra of the $Z^{+}_{c}$($\to \pi^+ J/\psi$) states from $\Upsilon(1S)$ decays
in signal MC samples. The $Z^{+}_{c}$ shapes are described by Breit-Wigner (BW) functions convolved with Gaussian functions.
The solid arrows show the required recoil-mass signal region.

To suppress the background level
effectively for $\Upsilon(1S)$ decays, we require the recoil mass of $\pi^{+}\jpsi$, $M_{\rm miss}(\pi^{+}\jpsi)$, to satisfy
$|M_{\rm miss}(\pi^{+}\jpsi)-m_{Z^{-}_{c}(3900)}|$ $<$ 0.03~GeV/$c^{2}$ for the
double-$Z_{c}(3900)$ mode or
$|M_{\rm miss}(\pi^{+}\jpsi)-m_{Z^{-}_{c}(4200)}|$ $<$ 0.21~GeV/$c^{2}$ for the
double-$Z_{c}(4200)$ and $Z_{c}(3900)$-plus-$Z_{c}(4200)$ modes,
where $m_{Z^{-}_{c}(3900)}$ and $m_{Z^{-}_{c}(4200)}$ are the nominal
masses~\cite{C38.090001}, and $M_{\rm miss}(\pi^{+}\jpsi)=\sqrt{( p_{e^+e^-}-p_{\pi^{+} \jpsi})^{2}}$ ($p_{e^+e^-}$ and $p_{\pi^+ \jpsi}$
are the four-momenta of the $e^+e^-$ and $\pi^+\jpsi$ systems).
These requirements maximize $S/\sqrt{S+B}$, where $S$ is the
number of fitted signal events in signal MC samples assuming
$\BR(\Upsilon(1S)\to Z^{+}_{c}Z^{(\prime) -}_{c})\times
\BR(Z^{+}_{c}\to \pi^{+}\jpsi)=10^{-5}$, and $B$ is
the number of estimated background events in the $Z^{+}_{c}$
signal region using inclusive MC samples.
The same requirements also maximize $S/\sqrt{S+B}$ for $\Upsilon(2S)$ decays and $e^+ e^-$
reactions at $\sqrt{s} =$ 10.52, 10.58, and 10.867~GeV.

\begin{figure*}[htbp]
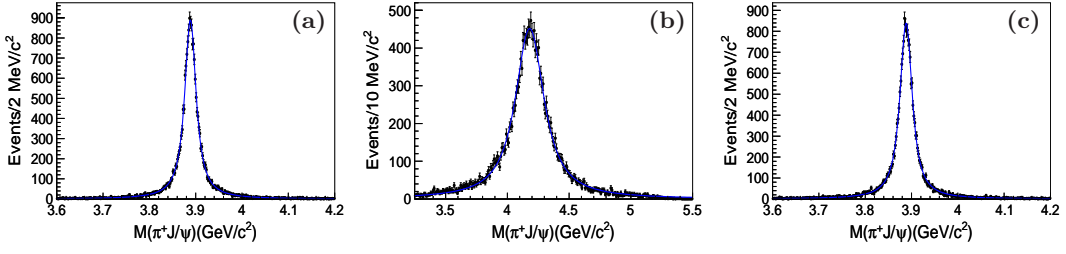

\includegraphics[height=4.5cm,width=3.2cm,angle=-90]{fig3a.epsi}
\put(-20, -10){\bf (a)}
\hspace{0.15cm}
\includegraphics[height=4.5cm,width=3.2cm,angle=-90]{fig3b.epsi}
\put(-20, -10){\bf (b)}
\hspace{0.15cm}
\includegraphics[height=4.5cm,width=3.2cm,angle=-90]{fig3c.epsi}
\put(-20, -10){\bf (c)}
\caption{The invariant mass distributions of the $Z^{+}_{c}$ ($\to \pi^+ J/\psi$) states from MC simulated (a) $\Upsilon(1S)\to Z^{+}_{c}(3900)Z^{-}_{c}(3900)$, (b) $\Upsilon(1S)\to Z^{+}_{c}(4200)Z^{-}_{c}(4200)$, and (c) $\Upsilon(1S)\to Z^{+}_{c}(3900)Z^{-}_{c}(4200)+c.c.$.}\label{M1}
\end{figure*}

After the application of the above requirements, Figs.~\ref{M1}(a)--(c) show the invariant mass distributions of the $Z^{+}_{c}$ states from $\Upsilon(1S)$ decays in the MC signal sample, where the solid lines show the fitted results with
a BW function convolved with a Gaussian function as $Z^{+}_{c}$ signal shapes.
Here, all combinations of $\pi^{+}\jpsi$ are retained, as is done for the modes $\pi^{+}\chi_{c1}$ and $\pi^{+}\psi(2S)$.
Fewer than 1\% of events have multiple entries here.
Based on the fitted results, the signal selection efficiencies are obtained and
listed in Table~\ref{T1}. Note that the efficiencies of the double-$Z_c(3900)$ and double-$Z_c(4200)$ modes are determined from the sum of the reconstructed $Z^{+}_{c}$ and $Z^{-}_{c}$
signals. This applies as well for double-$Z_c$ production in the other data sets with the same $Z_{c}$.

Figures~\ref{D11}(a)--(o) show the $\pi^+ J/\psi$ invariant-mass
spectra from the five data samples together with the
backgrounds from the normalized $\jpsi$ mass sideband events.
There are no evident signals for $Z^{+}_{c}$ states at the expected
positions. No peaking backgrounds are found in the $\jpsi$ mass
sideband events, nor in the $\Upsilon(1S,2S)$ and $e^{+}e^{-} \to
q\bar q$ ($q=u,~d,~s,~c$) inclusive MC samples.
There is a wide background enhancement at around 4.5~GeV/$c^{2}$ in the $\pi^+ J/\psi$ invariant-mass
distribution for $e^{+}e^{-} \to Z^{+}_{c}(4200)Z^{-}_{c}(4200)$ at $\sqrt{s} = 10.58$~GeV,
as shown in Fig.~\ref{D11}(k),
arising from $B$ decays to a $J/\psi$ meson.
Although these background events can be removed by requiring that the $J/\psi$ momentum in $e^+e^-$ CM
frame be larger than 2.5 GeV$/c$ ($P^{*}_{J/\psi} > 2.5$ GeV/$c$), we retain them since
the requirement of $P^{*}_{J/\psi} > 2.5$ GeV/$c$ decreases the signal efficiency by
40\% and this distinct feature is well described by a third-order Chebyshev polynomial function.

\begin{figure*}[htbp]
\includegraphics[width=3.3cm,angle=-90]{fig4a.epsi}
\hspace{0.15cm}
\includegraphics[width=3.3cm,angle=-90]{fig4b.epsi}
\hspace{0.15cm}
\vspace{0.35cm}
\includegraphics[width=3.3cm,angle=-90]{fig4c.epsi}
\hspace{0.15cm}
\includegraphics[width=3.3cm,angle=-90]{fig4d.epsi}
\hspace{0.15cm}
\includegraphics[width=3.3cm,angle=-90]{fig4e.epsi}
\hspace{0.15cm}
\vspace{0.35cm}
\includegraphics[width=3.3cm,angle=-90]{fig4f.epsi}
\hspace{0.15cm}
\includegraphics[width=3.3cm,angle=-90]{fig4g.epsi}
\hspace{0.15cm}
\includegraphics[width=3.3cm,angle=-90]{fig4h.epsi}
\hspace{0.15cm}
\vspace{0.35cm}
\includegraphics[width=3.3cm,angle=-90]{fig4i.epsi}
\hspace{0.15cm}
\includegraphics[width=3.3cm,angle=-90]{fig4j.epsi}
\hspace{0.15cm}
\includegraphics[width=3.3cm,angle=-90]{fig4k.epsi}
\hspace{0.15cm}
\vspace{0.35cm}
\includegraphics[width=3.3cm,angle=-90]{fig4l.epsi}
\hspace{0.15cm}
\includegraphics[width=3.3cm,angle=-90]{fig4m.epsi}
\hspace{0.15cm}
\includegraphics[width=3.3cm,angle=-90]{fig4n.epsi}
\hspace{0.15cm}
\includegraphics[width=3.3cm,angle=-90]{fig4o.epsi}
 \caption{The invariant mass spectra of the
$Z^{+}_{c}$ ($\to \pi^+ J/\psi$) states in the processes (a)
$\Upsilon(1S)\to Z^{+}_{c}(3900)Z^{-}_{c}(3900)$, (b)
$\Upsilon(1S)\to Z^{+}_{c}(4200)Z^{-}_{c}(4200)$, (c)
$\Upsilon(1S)\to Z^{+}_{c}(3900)Z^{-}_{c}(4200)+c.c.$; (d) $\Upsilon(2S)\to
Z^{+}_{c}(3900)Z^{-}_{c}(3900)$, (e) $\Upsilon(2S)\to
Z^{+}_{c}(4200)Z^{-}_{c}(4200)$, (f) $\Upsilon(2S)\to
Z^{+}_{c}(3900)Z^{-}_{c}(4200)+c.c.$; (g)-(i)
$e^{+}e^{-} \to Z^{+}_{c}(3900)Z^{-}_{c}(3900)$, $e^{+}e^{-}
\to Z^{+}_{c}(4200)Z^{-}_{c}(4200)$, and $e^{+}e^{-} \to
Z^{+}_{c}(3900)Z^{-}_{c}(4200)$+c.c. at $\sqrt{s}=10.52$~GeV; (j)-(l) $e^{+}e^{-} \to Z^{+}_{c}(3900)Z^{-}_{c}(3900)$,
$e^{+}e^{-} \to Z^{+}_{c}(4200)Z^{-}_{c}(4200)$, and $e^{+}e^{-}
\to Z^{+}_{c}(3900)Z^{-}_{c}(4200)+c.c.$ at $\sqrt{s}=10.58$~GeV; (m)-(o) $e^{+}e^{-} \to Z^{+}_{c}(3900)Z^{-}_{c}(3900)$, $e^{+}e^{-} \to Z^{+}_{c}(4200)Z^{-}_{c}(4200)$ and $e^{+}e^{-} \to Z^{+}_{c}(3900)Z^{-}_{c}(4200)+c.c.$ at $\sqrt{s}=10.867$~GeV. The
solid curves are the best fits described in the text, the dotted
lines are the fitted backgrounds, and the shaded histograms are
from the normalized $J/\psi$ mass sideband events.}\label{D11}
\end{figure*}

An unbinned extended maximum likelihood fit to each $\pi^+ J/\psi$
invariant-mass spectrum is performed to extract the signal and
background yields in the five data samples. The $Z^{+}_{c}$ signal shapes used in the fits are
BW functions convolved with a Gaussian function as
signal probability density functions (the parameters of BW function being fixed to the nominal masses and widths of $Z^{+}_{c}$ states and Gaussian functions being fixed to those from the fits to MC signal
samples with mass resolutions of 4 and 6 MeV/$c^2$ for $Z^{+}_{c}(3900)$ and $Z^{+}_{c}(4200)$, respectively). For the backgrounds, a second-order Chebyshev polynomial function is
adopted, except for $e^{+}e^{-} \to Z^{+}_{c}(4200)Z^{-}_{c}(4200)$ at $\sqrt{s} = 10.58$~GeV, where a third-order polynomial is used. The fitted results are shown in Fig.~\ref{D11} and summarized in Tables~\ref{T1} and \ref{T2}.

The $\BR(\Upsilon(1S,2S)\to
Z^{+}_{c}Z^{(\prime) -}_{c})\times\BR(Z^{+}_{c}\to\pi^{+}\jpsi)$ and
$\sigma(e^+e^- \to Z^{+}_{c}Z^{(\prime) -}_{c})\times \BR(Z^{+}_{c} \to
\pi^+ \jpsi)$ are calculated using
\begin{equation}
\frac{N_{\rm
fit}}{N_{\Upsilon(1S,2S)}\times\varepsilon\times\mathcal{B}_{\rm
decay}}
\end{equation}
and
\begin{equation}
\frac{N_{\rm fit}\times|1-\prod|^{2}}{\mathcal{L}
\times \mathcal{B}_{\rm decay} \times \varepsilon \times
(1+\delta)_{\rm ISR}},
\end{equation}
respectively, where $N_{\rm fit}$ is the
fitted $Z^{+}_{c}$ signal yield, $N_{\Upsilon(1S,2S)}$ is the
total number of $\Upsilon(1S,2S)$ events, $\varepsilon$ is the
corresponding selection efficiency, $\mathcal{L}$ is the
integrated luminosity, $\mathcal{B}_{\rm
decay}=\mathcal{B}(\jpsi\to\ell^+\ell^-)$~\cite{C38.090001},
$(1+\delta)_{\rm ISR}$ is the radiative correction factor and
$|1-\prod|^{2}$ is the vacuum polarization factor. The radiative
correction factors $(1+\delta)_{\rm ISR}$ are 0.650, 0.657, and
0.654 for $\sqrt{s} = 10.52$, 10.58, and 10.867 GeV, respectively, calculated using formulae given in Ref.~\cite{Fiz.41.733};
the corresponding values of $|1-\prod|^{2}$ are 0.931, 0.930, and
0.929~\cite{J.C.66.585}. In the calculation of $(1+\delta)_{\rm
ISR}$, we assume that the $s$ dependence of the cross section is $\sigma(e^{+}e^{-} \to A \bar{A}) \propto
1/s^2$.
The product branching fractions and the product of the Born
cross section and the branching fraction for the studied modes are
listed in Tables~\ref{T1} and ~\ref{T2}.

The statistical significances of the $Z^{+}_{c}$ signals are
calculated using
$\sqrt{-2\ln(\mathcal{L}_0/\mathcal{L}_{\rm max})}$
where $\mathcal{L}_0$ and $\mathcal{L}_{\rm max}$ are the
likelihoods of the fits without and with signal, respectively. The
values are summarized
in Tables~\ref{T1} and~\ref{T2}. Since the statistical
significance in each case is less than $3\sigma$, upper limits on
the signal yields ($N^{\rm UL}$), the product branching
fractions ($\BR^{\rm UL}(\Upsilon(1S,2S)\to
Z^{+}_{c}Z^{(\prime) -}_{c})\times\BR(Z^{+}_{c}\to\pi^{+}\jpsi)$), and the
product of the Born cross section and the branching fraction ($\sigma^{\rm
UL}(e^+e^- \to Z^{+}_{c}Z^{(\prime) -}_{c})\times \BR(Z^{+}_{c} \to
\pi^+ \jpsi)$), are determined at the $90\%$ credibility level
(C.L.)~\cite{C.L.} by solving the equation $\int^{x^{\rm UL}}_0
\mathcal{L}(x)dx / \int^{+\infty}_0\mathcal{L}(x)dx = 0.9$, where
$x$ is the assumed signal yield, product branching fraction, or
product of the Born cross section and the branching fraction, and
$\mathcal{L}(x)$ is the corresponding maximized likelihood for the
data. To take into account the systematic uncertainties discussed
in Sec.~\ref{secsys}, the likelihood is convolved with a Gaussian function whose
width equals to the corresponding systematic uncertainty.

The determined 90\% C.L. upper limits of $N^{\rm UL}$, $\BR^{\rm
UL}(\Upsilon(1S,2S)\to Z^{+}_{c}Z^{(\prime) -}_{c})\times\BR(Z^{+}_{c} \to
\pi^{+} \jpsi)$ and $\sigma^{\rm UL}(e^{+}e^{-} \to Z^{+}_{c}Z^{(\prime) -}_{c})\times\BR(Z^{+}_{c} \to \pi^{+} \jpsi)$ at $\sqrt{s} =$ 10.52, 10.58, and 10.867~GeV
are listed in Tables~\ref{T1} and~\ref{T2}, together with the signal yields ($N^{\rm fit}$), the
selection efficiencies ($\varepsilon$), the statistical
significances ($\Sigma$), the systematic uncertainties
($\sigma_{\rm syst}$) (discussed below), and the central values of
$\BR(\Upsilon(1S,2S)\to
Z^{+}_{c}Z^{(\prime) -}_{c})\times\BR(Z^{+}_{c}\to\pi^{+}\jpsi)$ and
$\sigma(e^+e^- \to Z^{+}_{c}Z^{(\prime) -}_{c})\times \BR(Z^{+}_{c} \to
\pi^+ \jpsi)$, with the total errors being the sum in quadrature of the statistical and systematic errors.

\begin{table*}[htbp]
\begin{threeparttable}
\caption{Summary of the 90\% C.L. upper limits on
$\BR(\Upsilon(1S,2S)\to
Z^{+}_{c}Z^{(\prime) -}_{c})\times\BR(Z^{+}_{c}\to\pi^{+}\jpsi)$ for
$Z^{+}_{c}(3900)Z^{-}_{c}(3900)$,
$Z^{+}_{c}(4200)Z^{-}_{c}(4200)$ and
$Z^{+}_{c}(3900)Z^{-}_{c}(4200)+c.c.$, where $N^{\rm fit}$ is the
signal yield,  $N^{\rm UL}$ is the 90\% C.L. upper
limit on the number of signal events, $\varepsilon$(\%) is the
selection efficiency, $\Sigma(\sigma)$ is the statistical signal
significance, $\sigma_{\rm syst}$(\%) is the total systematic
uncertainty, and $\BR$ and $\BR^{\rm UL}$ are the branching fraction
and the corresponding 90\% C.L. upper limit on the branching
fraction in units of $10^{-6}$.}\label{T1} \scriptsize
\begin{tabular}{c | c | c | c | c | c | c | c}
\hline \multirow{2}{*}{Mode} &\multirow{2}{*}{$N^{\rm fit}$}&\multirow{2}{*}{$N^{\rm UL}$}&$\varepsilon$&$\Sigma$&$\sigma_{\rm syst}$&$\BR(\Upsilon\to Z^{+}_{c}Z^{(\prime) -}_{c})\times$&$\BR^{\rm UL}(\Upsilon\to Z^{+}_{c}Z^{(\prime) -}_{c})\times$\\
&&&(\%)&$(\sigma)$&(\%)&$\BR(Z^{+}_{c}\to\pi^{+}\jpsi)$&$\BR(Z^{+}_{c}\to\pi^{+}\jpsi)$\\
\hline
$\Upsilon(1S)\to Z^{+}_{c}(3900)Z^{-}_{c}(3900)$&$0.9\pm4.3$  &9.2  &44.1  &0.2 &25.2&$0.2\pm0.7$ &$1.8$ \\
$\Upsilon(1S)\to Z^{+}_{c}(4200)Z^{-}_{c}(4200)$&$50.9\pm42.4$  &117.1  &44.6  &1.2&27.8&$9.4\pm8.3$  &$22.3$ \\
$\Upsilon(1S)\to Z^{+}_{c}(3900)Z^{-}_{c}(4200)+c.c.$&$3.0\pm10.1$  &22.0  &22.6  &0.3&20.5&$1.1\pm3.7$  &$8.1$ \\
$\Upsilon(2S)\to Z^{+}_{c}(3900)Z^{-}_{c}(3900)$&$-1.7\pm3.0$  &7.4  &41.1  &-&16.9&$-0.2\pm0.4$  &$1.0$ \\
$\Upsilon(2S)\to Z^{+}_{c}(4200)Z^{-}_{c}(4200)$&$58.0\pm47.9$  &125.2  &42.2  &1.2&31.4&$7.3\pm6.4$  &$16.7$ \\
$\Upsilon(2S)\to Z^{+}_{c}(3900)Z^{-}_{c}(4200)+c.c.$&$11.2\pm11.5$  &29.2 &21.3  &1.0&16.8&$2.8\pm2.9$  &$7.3$ \\
\hline
\end{tabular}
\end{threeparttable}
\end{table*}

\begin{table*}[htbp]
\begin{threeparttable}
\caption{Summary of the 90\% C.L. upper limits on $\sigma(e^+e^-
\to Z^{+}_{c}Z^{(\prime) -}_{c})\times \BR(Z^{+}_{c} \to \pi^+ \jpsi)$
for $Z^{+}_{c}(3900)Z^{-}_{c}(3900)$, $Z^{+}_{c}(4200)Z^{-}_{c}(4200)$ and $Z^{+}_{c}(3900)Z^{-}_{c}(4200)+c.c.$  at
$\sqrt{s}$ = 10.52, 10.58, and 10.867 GeV, where
$N^{\rm fit}$ is the signal yield, $N^{\rm UL}$ is the
90\% C.L. upper limit on the number of signal events,
$\varepsilon$(\%) is the selection efficiency, $\Sigma(\sigma)$ is
the statistical signal significance, $\sigma_{\rm syst}$(\%) is
the total systematic uncertainty, $\sigma$ is the Born cross
section $\sigma(e^+e^- \to Z^{+}_{c}Z^{(\prime) -}_{c})$, and
$\sigma^{\rm UL}$ is the corresponding 90\% C.L. upper limit in
units of fb.}\label{T2}
\scriptsize
\begin{tabular}{c | c | c | c | c | c | c | c | c }
\hline \multirow{2}{*}{Mode} &$\sqrt{s}$ & \multirow{2}{*}{$N^{\rm fit}$}&\multirow{2}{*}{$N^{\rm UL}$}&$\varepsilon$&$\Sigma$&$\sigma_{\rm syst}$&$\sigma\times$&$\sigma^{\rm UL}\times$\\
&(GeV)&&&(\%)&($\sigma$)&(\%)&$\BR(Z^{+}_{c}\to\pi^{+}\jpsi)$&$\BR(Z^{+}_{c}\to\pi^{+}\jpsi)$\\
\hline
$e^{+}e^{-} \to Z^{+}_{c}(3900)Z^{-}_{c}(3900)$ & $10.52$ &$-4.9\pm3.6$  &7.2  &41.5  &-&10.3&$-1.6\pm1.2$ &$2.3$ \\
$e^{+}e^{-} \to Z^{+}_{c}(4200)Z^{-}_{c}(4200)$ & $10.52$ &$-27.5\pm57.8$  &82.8  &43.7  &-&34.2&$-8.5\pm18.1$ &$26.5$ \\
$e^{+}e^{-} \to Z^{+}_{c}(3900)Z^{-}_{c}(4200)+c.c.$ & $10.52$ &$-0.5\pm15.0$  &28.4  &21.0  &-&22.9&$-0.3\pm9.7$ &$18.3$ \\
$e^{+}e^{-} \to Z^{+}_{c}(3900)Z^{-}_{c}(3900)$ & $10.58$ &$11.8\pm13.0$  &32.2  &41.5  &0.9&12.7&$0.5\pm0.5$ &$1.3$ \\
$e^{+}e^{-} \to Z^{+}_{c}(4200)Z^{-}_{c}(4200)$ & $10.58$ &$132.1\pm173.0$  &390.1  &43.4  &0.8&35.4&$5.1\pm6.9$ &$15.5$ \\
$e^{+}e^{-} \to Z^{+}_{c}(3900)Z^{-}_{c}(4200)+c.c.$ & $10.58$ &$-7.7\pm39.4$  &63.4  &20.8  &-&20.7&$-0.6\pm3.2$ &$5.1$ \\
$e^{+}e^{-} \to Z^{+}_{c}(3900)Z^{-}_{c}(3900)$ & $10.867$ &$-1.4\pm4.6$  &9.0  &41.5  &-&17.0&$-0.3\pm1.1$ &$2.2$ \\
$e^{+}e^{-} \to Z^{+}_{c}(4200)Z^{-}_{c}(4200)$ & $10.867$ &$-0.2\pm41.6$  &93.7  &43.7  &-&33.2&$-0.1\pm9.4$ &$21.9$ \\
$e^{+}e^{-} \to Z^{+}_{c}(3900)Z^{-}_{c}(4200)+c.c.$ & $10.867$ &$30.3\pm16.7$  &53.9  &20.5  &1.9&16.3&$14.6\pm8.4$ &$26.6$ \\
\hline
\end{tabular}
\end{threeparttable}
\end{table*}

\section{\boldmath Search for $\Upsilon(1S,2S) \to Z^{+}_{c}Z^{(\prime) -}_{c}$
and $e^{+}e^{-} \to Z^{+}_{c}Z^{(\prime) -}_{c}$ at $\sqrt{s} =$ 10.52, 10.58, and 10.867~GeV
with $Z^{+}_{c} \to \pi^{+}\chi_{c1}(1P), \pi^+\psi(2S)$}

In the five data sets, we search for the production of double-$Z_{c1}(4050)$ states, double-$Z_{c2}(4250)$ states and $Z_{c1}(4050)$-plus-$Z_{c2}(4250)$ states, where one
of the $Z_c$ decays into $\pi^{+}$ and $\chi_{c1}(1P)$ and the other decays inclusively; and double-$Z_{c}(4050)$ states, double-$Z_{c}(4430)$ states and $Z_{c}(4050)$-plus-$Z_{c}(4430)$ states, where one
of the $Z_c$ decays into $\pi^{+}$ and $\psi(2S)$ and
the other decays inclusively. The $\chi_{c1}(1P)$ and $\psi(2S)$ are
reconstructed via their decays into $\gamma \jpsi$ and $\pi^{+}
\pi^{-} \jpsi$, respectively, with $\jpsi \to \ell^+\ell^-$.

\begin{figure*}[htbp]
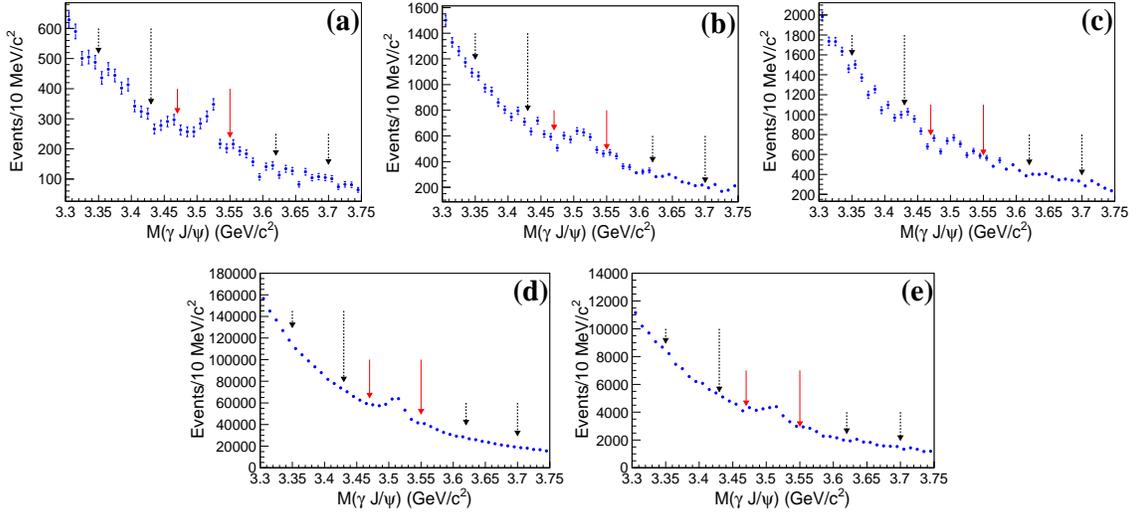

\includegraphics[height=4.9cm,width=3.2cm,angle=-90]{fig5a.epsi}
\includegraphics[height=4.9cm,width=3.2cm,angle=-90]{fig5b.epsi}
\vspace{0.3cm}
\includegraphics[height=4.9cm,width=3.2cm,angle=-90]{fig5c.epsi}
\includegraphics[height=5cm,width=3.2cm,angle=-90]{fig5d.epsi}
\includegraphics[height=5cm,width=3.2cm,angle=-90]{fig5e.epsi}
\caption{The invariant mass distributions of the $\chi_{c1}(1P)$
candidates  from (a) $\Upsilon(1S)$, (b) $\Upsilon(2S)$, (c)
$\sqrt{s} = 10.52$~GeV, (d) $\sqrt{s} = 10.58$~GeV and (e)
$\sqrt{s} = 10.867$~GeV data samples. The solid red arrows show the
$\chi_{c1}(1P)$  signal region and the dashed black arrows show the
$\chi_{c1}(1P)$ mass sideband regions.}\label{mchic1data}
\end{figure*}

After requiring the mass of the lepton pair to be within the $\jpsi$ signal
region ($|M_{\ell^+\ell^-}-m_{\jpsi}|<0.03$~GeV/$c^2$),
Figs.~\ref{mchic1data}(a)--(e) and \ref{mpsi2Sdata}(a)--(e) show
the invariant mass distributions of the $\chi_{c1}(1P)$ and $\psi(2S)$
candidates from five data samples. Clear $\chi_{c1}(1P)$ signals could be seen in $\Upsilon(1S)$ decay and in $e^{+}e^{-}$ annihilation at $\sqrt{s} =$10.58 and 10.867~GeV, and evidence for $\chi_{c1}(1P)$ is observed in $\Upsilon(2S)$ decay;
clear $\psi(2S)$ signals are observed in $\Upsilon(1S,2S)$ decay and in $e^{+}e^{-}$ annihilation at $\sqrt{s} =$ 10.52, 10.58
and 10.867~GeV. We define the $\chi_{c1}(1P)$ and
$\psi(2S)$ signal and sideband regions as follows. The $\chi_{c1}(1P)$
signal region is $|M_{\gamma\jpsi}-m_{\chi_{c1}(1P)}|<0.04$~GeV/$c^2$
($>98\%$ signal events are reserved according to the MC simulation), where $m_{\chi_{c1}(1P)}$ is the nominal mass~\cite{C38.090001}, while the $\chi_{c1}(1P)$
mass sidebands are $3.35<M_{\gamma\jpsi}<3.43$~GeV/$c^2$ or
$3.62<M_{\gamma\jpsi}<3.70$~GeV/$c^2$ (twice the width of the signal region). The $\psi(2S)$ signal region is $|M_{\pi^{+}
\pi^{-} \jpsi}-m_{\psi(2S)}|<0.007$~GeV/$c^2$ ($>98\%$ signal events are reserved according to the MC simulation), where $m_{\psi(2S)}$ is the nominal
mass~\cite{C38.090001}, while the $\psi(2S)$ mass sidebands are
$3.65<M_{\pi^{+} \pi^{-} \jpsi}<3.664$~GeV/$c^2$ or
$3.71<M_{\pi^{+} \pi^{-} \jpsi}<3.724$~GeV/$c^2$ (twice the width of the signal region).

\begin{figure*}[htbp]
\includegraphics[height=4.9cm,width=3.2cm,angle=-90]{fig6a.epsi}
\includegraphics[height=4.9cm,width=3.2cm,angle=-90]{fig6b.epsi}
\vspace{0.3cm}
\includegraphics[height=4.9cm,width=3.2cm,angle=-90]{fig6c.epsi}
\includegraphics[height=5cm,width=3.2cm,angle=-90]{fig6d.epsi}
\includegraphics[height=5cm,width=3.2cm,angle=-90]{fig6e.epsi}
\caption{The invariant mass distributions of the $\psi(2S)$
candidates from (a) $\Upsilon(1S)$, (b) $\Upsilon(2S)$, (c)
$\sqrt{s} = 10.52$~GeV, (d) $\sqrt{s} = 10.58$~GeV and (e)
$\sqrt{s} = 10.867$~GeV data samples. The solid red arrows show the
$\psi(2S)$ signal region and the dashed black arrows show the $\psi(2S)$
mass sideband regions.}\label{mpsi2Sdata}
\end{figure*}

As before, for $\Upsilon(1S)$ decays, we optimize the requirements of recoil mass of
$\pi^{+}\chi_{c1}(1P)$  and $\pi^{+}\psi(2S)$ by maximizing
$S/\sqrt{S+B}$. The optimized requirements are
$|M_{\rm miss}(\pi^{+}\chi_{c1}(1P))-m_{Z^{-}_{c1}(4050)}|$ $<$ 0.08~GeV/$c^{2}$
for the $Z^{+}_{c1}(4050)$$Z^{-}_{c1}(4050)$ mode and
$|M_{\rm miss}(\pi^{+}\chi_{c1}(1P))-m_{Z^{-}_{c2}(4250)}|$ $<$ 0.13~GeV/$c^{2}$ for the
$Z^{+}_{c2}(4250)$$Z^{-}_{c2}(4250)$ and $Z^{+}_{c1}(4050)$$Z^{-}_{c2}(4250)+c.c.$ modes; they are
$|M_{\rm miss}(\pi^{+}\psi(2S))-m_{Z^{-}_{c}(4050)}|$ $<$ 0.06~GeV/$c^{2}$
for the $Z^{+}_{c}(4050)$$Z^{-}_{c}(4050)$ mode and
$|M_{\rm miss}(\pi^{+}\psi(2S))-m_{Z^{-}_{c}(4430)}|$ $<$ 0.12~GeV for the
$Z^{+}_{c}(4430)$$Z^{-}_{c}(4430)$ and $Z^{+}_{c}(4050)$$Z^{-}_{c}(4430)+c.c.$ modes.
For $\Upsilon(2S)$ decays and $e^+ e^-$ reactions at $\sqrt{s} =$ 10.52, 10.58, and 10.867~GeV,
the optimized requirements are the same as $\Upsilon(1S)$.

\begin{figure*}[htbp]
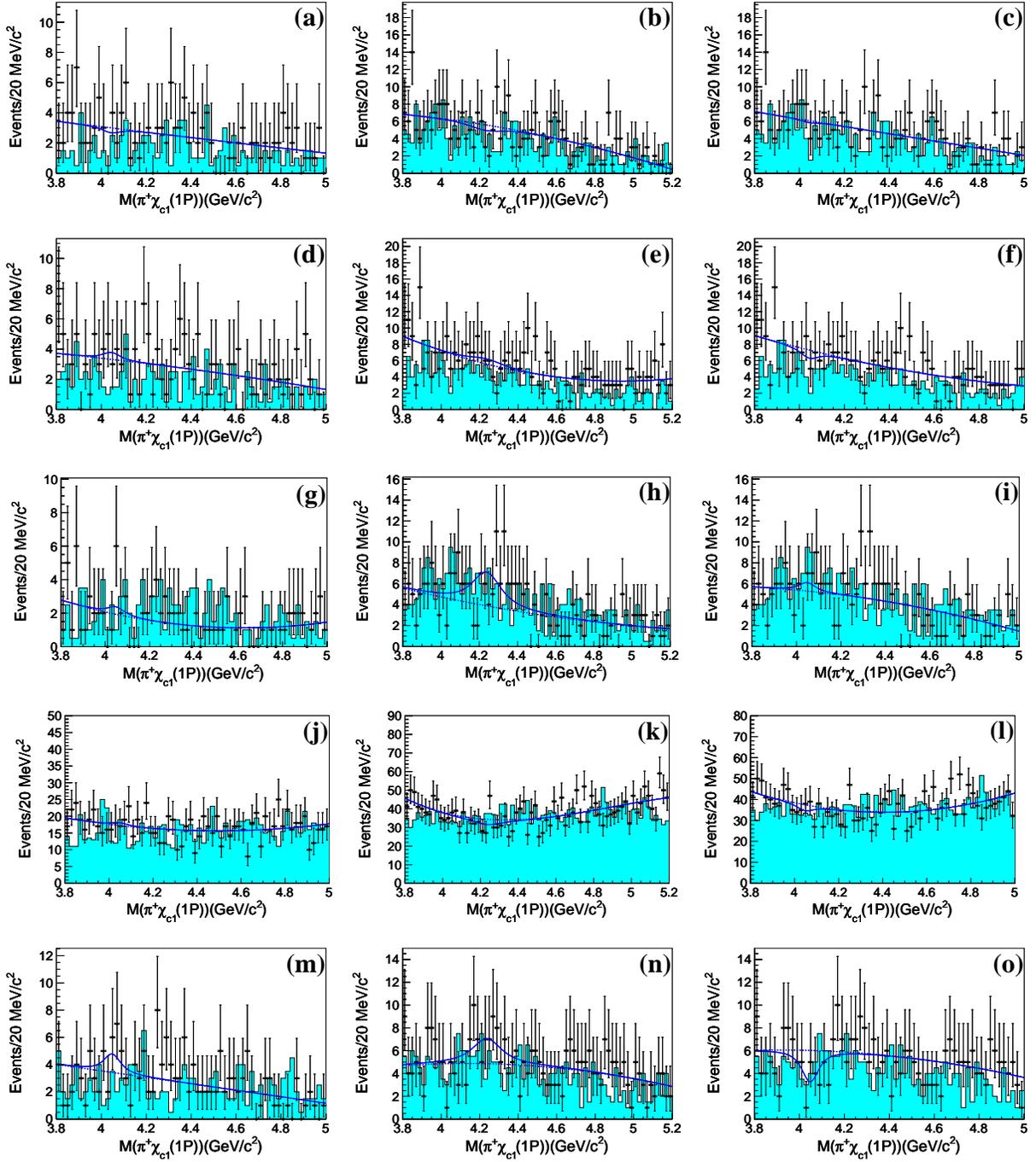

\includegraphics[width=3.3cm,angle=-90]{fig7a.epsi}
\hspace{0.15cm}
\includegraphics[width=3.3cm,angle=-90]{fig7b.epsi}
\hspace{0.15cm}
\vspace{0.35cm}
\includegraphics[width=3.3cm,angle=-90]{fig7c.epsi}
\hspace{0.15cm}
\includegraphics[width=3.3cm,angle=-90]{fig7d.epsi}
\hspace{0.15cm}
\includegraphics[width=3.3cm,angle=-90]{fig7e.epsi}
\hspace{0.15cm}
\vspace{0.35cm}
\includegraphics[width=3.3cm,angle=-90]{fig7f.epsi}
\hspace{0.15cm}
\includegraphics[width=3.3cm,angle=-90]{fig7g.epsi}
\hspace{0.15cm}
\includegraphics[width=3.3cm,angle=-90]{fig7h.epsi}
\hspace{0.15cm}
\vspace{0.35cm}
\includegraphics[width=3.3cm,angle=-90]{fig7i.epsi}
\hspace{0.15cm}
\includegraphics[width=3.3cm,angle=-90]{fig7j.epsi}
\hspace{0.15cm}
\includegraphics[width=3.3cm,angle=-90]{fig7k.epsi}
\hspace{0.15cm}
\vspace{0.35cm}
\includegraphics[width=3.3cm,angle=-90]{fig7l.epsi}
\hspace{0.15cm}
\includegraphics[width=3.3cm,angle=-90]{fig7m.epsi}
\hspace{0.15cm}
\includegraphics[width=3.3cm,angle=-90]{fig7n.epsi}
\hspace{0.15cm}
\includegraphics[width=3.3cm,angle=-90]{fig7o.epsi}
\caption{The invariant mass spectra of the $Z^{+}_{c}$ ($\to \pi^+
\chi_{c1}(1P)$) states in the processes (a) $\Upsilon(1S)\to
Z^{+}_{c1}(4050)Z^{-}_{c1}(4050)$, (b) $\Upsilon(1S)\to
Z^{+}_{c2}(4250)Z^{-}_{c2}(4250)$, (c) $\Upsilon(1S)\to
Z^{+}_{c1}(4050)Z^{-}_{c2}(4250)+c.c.$; (d)
$\Upsilon(2S)\to Z^{+}_{c1}(4050)Z^{-}_{c1}(4050)$, (e)
$\Upsilon(2S)\to Z^{+}_{c2}(4250)Z^{-}_{c2}(4250)$, (f)
$\Upsilon(2S)\to Z^{+}_{c1}(4050)Z^{-}_{c2}(4250)+c.c.$; (g)-(i) $e^{+}e^{-} \to Z^{+}_{c1}(4050)
Z^{-}_{c1}(4050)$, $e^{+}e^{-} \to Z^{+}_{c2}(4250)
Z^{-}_{c2}(4250)$, and $e^{+}e^{-} \to Z^{+}_{c1}(4050)
Z^{-}_{c2}(4250)+c.c.$ at $\sqrt{s}=10.52$~GeV; (j)-(l)
$e^{+}e^{-} \to Z^{+}_{c1}(4050)Z^{-}_{c1}(4050)$,
$e^{+}e^{-} \to Z^{+}_{c2}(4250)Z^{-}_{c2}(4250)$, and
$e^{+}e^{-} \to Z^{+}_{c1}(4050)Z^{-}_{c2}(4250)+c.c.$ at
$\sqrt{s}=10.58$~GeV; (m)-(o) $e^{+}e^{-} \to
Z^{+}_{c1}(4050)Z^{-}_{c1}(4050)$, $e^{+}e^{-} \to
Z^{+}_{c2}(4250)Z^{-}_{c2}(4250)$ and $e^{+}e^{-} \to
Z^{+}_{c1}(4050)Z^{-}_{c2}(4250)$ at $\sqrt{s}=10.867$~GeV. The solid curves are the best fits described in the
text, the dotted lines are the fitted backgrounds, and the shaded
histograms are from the normalized $\chi_{c1}(1P)$ mass sideband
events.}\label{D21}
\end{figure*}

After the application of these requirements, Figs.~\ref{D21}(a)--(o) and
\ref{D31}(a)--(o) show the $\pi^{+}\chi_{c1}(1P)$ and
$\pi^{+}\psi(2S)$ invariant mass spectra from the five data samples, together
with the backgrounds from the normalized $\chi_{c1}(1P)$ or $\psi(2S)$
mass sideband events. There are no evident signals for $Z^{+}_{c}$
states at the expected position. No peaking backgrounds are found in
the $\chi_{c1}(1P)$ or $\psi(2S)$ mass sideband events, nor in the
$\Upsilon(1S,2S)$ and $e^{+}e^{-} \to q\bar q$ inclusive MC
samples.

\begin{figure*}[htbp]
\includegraphics[width=3.3cm,angle=-90]{fig8a.epsi}
\hspace{0.15cm}
\includegraphics[width=3.3cm,angle=-90]{fig8b.epsi}
\hspace{0.15cm}
\vspace{0.35cm}
\includegraphics[width=3.3cm,angle=-90]{fig8c.epsi}
\hspace{0.15cm}
\includegraphics[width=3.3cm,angle=-90]{fig8d.epsi}
\hspace{0.15cm}
\includegraphics[width=3.3cm,angle=-90]{fig8e.epsi}
\hspace{0.15cm}
\vspace{0.35cm}
\includegraphics[width=3.3cm,angle=-90]{fig8f.epsi}
\hspace{0.15cm}
\includegraphics[width=3.3cm,angle=-90]{fig8g.epsi}
\hspace{0.15cm}
\includegraphics[width=3.3cm,angle=-90]{fig8h.epsi}
\hspace{0.15cm}
\vspace{0.35cm}
\includegraphics[width=3.3cm,angle=-90]{fig8i.epsi}
\hspace{0.15cm}
\includegraphics[width=3.3cm,angle=-90]{fig8j.epsi}
\hspace{0.15cm}
\includegraphics[width=3.3cm,angle=-90]{fig8k.epsi}
\hspace{0.15cm}
\vspace{0.35cm}
\includegraphics[width=3.3cm,angle=-90]{fig8l.epsi}
\hspace{0.15cm}
\includegraphics[width=3.3cm,angle=-90]{fig8m.epsi}
\hspace{0.15cm}
\includegraphics[width=3.3cm,angle=-90]{fig8n.epsi}
\hspace{0.15cm}
\includegraphics[width=3.3cm,angle=-90]{fig8o.epsi}
\caption{The invariant mass spectra of the $Z^{+}_{c}$ ($\to \pi^+
\psi(2S)$) states in the processes (a) $\Upsilon(1S)\to
Z^{+}_{c}(4050)Z^{-}_{c}(4050)$, (b) $\Upsilon(1S)\to
Z^{+}_{c}(4430)Z^{-}_{c}(4430)$, (c) $\Upsilon(1S)\to
Z^{+}_{c}(4050)Z^{-}_{c}(4430)+c.c.$; (d)
$\Upsilon(2S)\to Z^{+}_{c}(4050)Z^{-}_{c}(4050)$, (e)
$\Upsilon(2S)\to Z^{+}_{c}(4430)Z^{-}_{c}(4430)$, (f)
$\Upsilon(2S)\to Z^{+}_{c}(4050)Z^{-}_{c}(4430)+c.c.$; (g)-(i) $e^{+}e^{-} \to Z^{+}_{c}(4050)
Z^{-}_{c}(4050)$, $e^{+}e^{-} \to Z^{+}_{c}(4430)
Z^{-}_{c}(4430)$, and $e^{+}e^{-} \to Z^{+}_{c}(4050)
Z^{-}_{c}(4430)+c.c.$ at $\sqrt{s}=10.52$~GeV; (j)-(l)
$e^{+}e^{-} \to Z^{+}_{c}(4050)Z^{-}_{c}(4050)$, $e^{+}e^{-}
\to Z^{+}_{c}(4430)Z^{-}_{c}(4430)$, and $e^{+}e^{-} \to
Z^{+}_{c}(4050)Z^{-}_{c}(4430)+c.c.$ at $\sqrt{s}=10.58$~GeV; (m)-(o) $e^{+}e^{-} \to Z^{+}_{c}(4050)Z^{-}_{c}(4050)$,
$e^{+}e^{-} \to Z^{+}_{c}(4430)Z^{-}_{c}(4430)$ and
$e^{+}e^{-} \to Z^{+}_{c}(4050)Z^{-}_{c}(4430)+c.c.$ at
$\sqrt{s}=10.867$~GeV. The solid curves are the best
fits described in the text, the dotted lines are the fitted
backgrounds, and the shaded histograms are from the normalized
$\psi(2S)$ mass sideband events.}\label{D31}
\end{figure*}

An unbinned extended maximum likelihood fit to the
$\pi^+\chi_{c1}(1P)$ ($\pi^+\psi(2S)$) invariant mass spectra is
performed to extract the signal and background yields from
$\Upsilon(1S,2S)$ decays and $e^{+}e^{-} \to Z^{+}_{c}Z^{(\prime) -}_{c}$ reactions at $\sqrt{s}$ = 10.52, 10.58, and 10.867 GeV.
The $Z^{+}_{c}$ signal probability density functions used in the fits are the convolutions of a BW
function with a Gaussian function (the parameters of BW function being fixed to the nominal masses and widths of $Z^{+}_{c}$ states and Gaussian functions being fixed to those from the fits to signal MC samples
with mass resolutions of 15, 16, 10, and 11 MeV/$c^{2}$ for
$Z^{+}_{c1}(4050)$, $Z^{+}_{c2}(4250)$, $Z^{+}_{c}(4050)$, and $Z^{+}_{c}(4250)$,
respectively). For the
backgrounds, a second-order Chebyshev polynomial function is
adopted. The fitted results are shown in Figs.~\ref{D21} and
\ref{D31} and summarized in Tables~\ref{T3} and \ref{T4}.

Marginal signals are seen in each fit. The largest statistical signal significance is 2.6$\sigma$ from  $e^{+}e^{-} \to Z_{c2}(4250)+Z_{c2}(4250)$ at $\sqrt{s} = 10.52$~GeV.
We use an ensemble of simulated experiments to estimate the
probability that background fluctuations alone would produce signals
as significant as those seen in the data. We generate $\pi^+\chi_{c1}(1P)$
mass spectra based on a fitted second-order Chebyshev polynomial function
alone with the same observed events in data, and search for the most
significant fluctuation in each mass spectrum in the studied mass range. From these pure-background
spectra, we obtain the distribution for $-2\ln(\mathcal{L}_0/\mathcal{L}_{\rm max})$, and compare it with the signal in the data.
In a total of 10,000 simulations, we find 1373 trials
with a $-2\ln(\mathcal{L}_0/\mathcal{L}_{\rm max})$ value greater than or equal to the
value obtained in the data. The resulting $p$ value is 0.1373,
corresponding to a significance of $1.5\sigma$.

Using the same method as described in the previous section, we
determine the 90\% C.L. upper limits on $\BR(\Upsilon(1S,2S)\to
Z^{+}_{c}Z^{(\prime) -}_{c})\times\BR(Z^{+}_{c} \to \pi^{+}
\chi_{c1}(1P)/\pi^+\psi(2S))$ and $\sigma(e^{+}e^{-} \to Z^{+}_{c}
Z^{(\prime) -}_{c})\times\BR(Z^{+}_{c} \to \pi^{+} \chi_{c1}(1P)/\pi^+
\psi(2S))$ at $\sqrt{s} =$ 10.52, 10.58, and 10.867~GeV. The values are listed in Tables~\ref{T3} and
\ref{T4}, together with the signal yields ($N^{\rm fit}$), the
90\% C.L. upper limits on the numbers of signal events ($N^{\rm
UL}$), the selection efficiencies ($\varepsilon$), the signal
significances ($\Sigma$), the systematic uncertainties
($\sigma_{\rm syst}$) discussed below, and the central values of
$\BR(\Upsilon(1S,2S)\to Z^{+}_{c}Z^{(\prime) -}_{c})\times
\BR(Z^{+}_{c}\to\pi^{+}\chi_{c1}(1P)/\pi^+\psi(2S))$ and
$\sigma(e^+e^- \to Z^{+}_{c}Z^{(\prime) -}_{c})\times \BR(Z^{+}_{c} \to
\pi^+ \chi_{c1}(1P)/\pi^+\psi(2S))$, with the total errors being the sum in quadrature of the statistical and systematic errors.

\begin{table*}[htbp]
\begin{threeparttable}
\caption{Summary of the 90\% C.L. upper limits on
$\BR(\Upsilon(1S,2S)\to
Z^{+}_{c}Z^{(\prime) -}_{c})\times\BR(Z^{+}_{c}\to\pi^{+}\chi_{c1}(1P)/\pi^+
\psi(2S))$ for $Z^{+}_{c1}(4050)Z^{-}_{c1}(4050)$,
$Z^{+}_{c2}(4250)Z^{-}_{c2}(4250)$ and
$Z^{+}_{c1}(4050)Z^{-}_{c2}(4250)+c.c.$ to $\pi^{+}\chi_{c1}(1P) +
anything$; and $Z^{+}_{c}(4050)Z^{-}_{c}(4050)$,
$Z^{+}_{c}(4430)Z^{-}_{c}(4430)$ and
$Z^{+}_{c}(4050)Z^{-}_{c}(4430)+c.c.$ to $\pi^{+}\psi(2S) + anything$,
where $N^{\rm fit}$ is the signal yield, $N^{\rm UL}$
is the 90\% C.L. upper limit on the number of signal events,
$\varepsilon$(\%) is the selection efficiency, $\Sigma(\sigma)$ is
the statistical signal significance, $\sigma_{\rm syst}$(\%) is
the total systematic uncertainty, and $\BR$ and $\BR^{\rm UL}$ are the
branching fraction and the corresponding 90\% C.L. upper limit on
the branching fraction in units of $10^{-6}$.}\label{T3}
\scriptsize
\begin{tabular}{c | c | c | c | c | c | c | c }
\hline \multirow{2}{*}{Mode} &\multirow{2}{*}{$N^{\rm fit}$}&\multirow{2}{*}{$N^{\rm UL}$}&$\varepsilon$&$\Sigma$&$\sigma_{\rm syst}$&$\BR(\Upsilon\to Z^{+}_{c}Z^{(\prime) -}_{c})\times$&$\BR^{\rm UL}(\Upsilon\to Z^{+}_{c}Z^{(\prime) -}_{c})\times$\\
&&&(\%)&($\sigma$)&(\%)&$\BR(Z^{+}_{c}\to\pi^{+}\chi_{c1}(1P)/\pi^+ \psi(2S))$&$\BR(Z^{+}_{c}\to\pi^{+}\chi_{c1}(1P)/\pi^+ \psi(2S))$\\
\hline
$\Upsilon(1S)\to Z^{+}_{c1}(4050)Z^{-}_{c1}(4050)$&$-2.1\pm7.2$  &13.1  &21.6  &-&41.3&$-2.4\pm8.1$ &$15.8$ \\
$\Upsilon(1S)\to Z^{+}_{c2}(4250)Z^{-}_{c2}(4250)$&$-8.3\pm14.1$  &21.7  &20.9  &-&34.5&$-9.7\pm16.8$  &$26.6$ \\
$\Upsilon(1S)\to Z^{+}_{c1}(4050)Z^{-}_{c2}(4250)+c.c.$&$-1.3\pm10.2$  &18.1  &10.1  &-&25.9&$-3.1\pm24.3$  &$44.2$ \\
$\Upsilon(2S)\to Z^{+}_{c1}(4050)Z^{-}_{c1}(4050)$&$2.9\pm7.7$  &16.4  &20.5  &0.2&38.7&$2.2\pm5.9$  &$13.5$ \\
$\Upsilon(2S)\to Z^{+}_{c2}(4250)Z^{-}_{c2}(4250)$&$8.1\pm15.8$  &32.6  &19.2  &0.5&34.0&$6.6\pm13.2$  &$26.7$ \\
$\Upsilon(2S)\to Z^{+}_{c1}(4050)Z^{-}_{c2}(4250)+c.c.$&$-6.3\pm10.5$  &16.1 &9.4  &-&18.0&$-10.5\pm17.6$  &$27.2$ \\\hline
$\Upsilon(1S)\to Z^{+}_{c}(4050)Z^{-}_{c}(4050)$&$6.7\pm5.3$  &14.4  &16.0  &1.4&16.4&$10.0\pm8.1$ &$23.3$ \\
$\Upsilon(1S)\to Z^{+}_{c}(4430)Z^{-}_{c}(4430)$&$-1.5\pm7.6$  &13.6  &16.2  &-&22.0&$-2.2\pm11.3$  &$20.3$ \\
$\Upsilon(1S)\to Z^{+}_{c}(4050)Z^{-}_{c}(4430)+c.c.$&$4.4\pm5.7$  &14.3  &7.7  &0.8&28.8&$13.6\pm18.1$  &$45.5$ \\
$\Upsilon(2S)\to Z^{+}_{c}(4050)Z^{-}_{c}(4050)$&$-1.9\pm6.4$  &10.8  &15.1  &-&16.1&$-1.9\pm6.6$  &$11.1$ \\
$\Upsilon(2S)\to Z^{+}_{c}(4430)Z^{-}_{c}(4430)$&$3.4\pm9.6$  &20.0  &15.3  &0.3&17.8&$3.4\pm9.7$  &$20.3$ \\
$\Upsilon(2S)\to Z^{+}_{c}(4050)Z^{-}_{c}(4430)+c.c.$&$-4.9\pm6.2$  &10.2 &7.5  &-&26.1&$-10.1\pm13.1$  &$21.1$ \\
\hline
\end{tabular}
\end{threeparttable}
\end{table*}

\begin{table*}[htbp]
\begin{threeparttable}
\caption{Summary of the 90\% C.L. upper limits on $\sigma(e^+e^-
\to Z^{+}_{c}Z^{(\prime) -}_{c})\times \BR(Z^{+}_{c} \to \pi^+
\chi_{c1}(1P)/\pi^{+}\psi(2S))$ for
$Z^{+}_{c1}(4050)Z^{-}_{c1}(4050)$,
$Z^{+}_{c2}(4250)Z^{-}_{c2}(4250)$ and
$Z^{+}_{c1}(4050)Z^{-}_{c2}(4250)+c.c.$ to $\pi^{+}\chi_{c1}(1P) +
anything$; and $Z^{+}_{c}(4050)Z^{-}_{c}(4050)$,
$Z^{+}_{c}(4430)Z^{-}_{c}(4430)$ and
$Z^{+}_{c}(4050)Z^{-}_{c}(4430)+c.c.$ to $\pi^{+}\psi(2S) + anything$
at $\sqrt{s}$ = 10.52, $10.58$, and $10.867$~GeV where $N^{\rm fit}$ is the signal yield, $N^{\rm UL}$
is the 90\% C.L. upper limit on the number of signal events,
$\varepsilon$(\%) is the selection efficiency, $\Sigma(\sigma)$ is
the statistical signal significance, $\sigma_{\rm syst}$(\%) is
the total systematic uncertainty, $\sigma$ is the Born cross
section $\sigma(e^+e^- \to Z^{+}_{c}Z^{(\prime) -}_{c})$, and
$\sigma^{\rm UL}$ is the corresponding 90\% C.L. upper limit in
units of fb.}\label{T4} \scriptsize
\begin{tabular}{c | c | c | c | c | c | c | c | c}
\hline \multirow{2}{*}{Mode} &$\sqrt{s}$&\multirow{2}{*}{$N^{\rm fit}$}&\multirow{2}{*}{$N^{\rm UL}$}&$\varepsilon$&$\Sigma$&$\sigma_{\rm syst}$&$\sigma\times\BR(Z^{+}_{c}$&$\sigma^{\rm UL}\times\BR(Z^{+}_{c}$\\
&(GeV)&&&(\%)&($\sigma$)&(\%)&$\to\pi^{+}\chi_{c1}(1P)/\pi^{+}\psi(2S))$&$\to\pi^{+}\chi_{c1}(1P)/\pi^{+}\psi(2S))$\\
\hline
$e^{+}e^{-} \to Z^{+}_{c1}(4050)Z^{-}_{c1}(4050)$ & $10.52$&$1.2\pm6.5$  &13.2  &20.9  &0.2&28.3&$2.3\pm12.4$ &$25.0$ \\
$e^{+}e^{-} \to Z^{+}_{c2}(4250)Z^{-}_{c2}(4250)$ & $10.52$&$40.9\pm16.8$  &65.1  &19.4  &2.6&32.9&$83.9\pm44.1$ &$143.9$ \\
$e^{+}e^{-} \to Z^{+}_{c1}(4050)Z^{-}_{c2}(4250)+c.c.$ & $10.52$&$5.2\pm10.4$  &21.5  &9.5  &0.5&33.0&$21.7\pm44.1$ &$93.2$ \\
$e^{+}e^{-} \to Z^{+}_{c1}(4050)Z^{-}_{c1}(4050)$ & $10.58$&$4.1\pm18.9$  &36.3  &20.5  &0.2&21.9&$1.0\pm4.6$ &$8.8$ \\
$e^{+}e^{-} \to Z^{+}_{c2}(4250)Z^{-}_{c2}(4250)$ & $10.58$&$-35.2\pm48.3$  &25.7  &19.2  &-&45.8&$-9.0\pm13.1$ &$7.1$ \\
$e^{+}e^{-} \to Z^{+}_{c1}(4050)Z^{-}_{c2}(4250)+c.c.$ & $10.58$&$-18.0\pm24.8$  &34.5  &9.8  &-&45.0&$-9.1\pm13.2$ &$18.2$ \\
$e^{+}e^{-} \to Z^{+}_{c1}(4050)Z^{-}_{c1}(4050)$ & $10.867$&$8.6\pm8.5$  &23.0  &19.4  &1.0&26.0&$12.9\pm13.2$ &$35.7$ \\
$e^{+}e^{-} \to Z^{+}_{c2}(4250)Z^{-}_{c2}(4250)$ & $10.867$&$27.7\pm16.1$  &49.5  &18.5  &1.7&27.0&$43.6\pm28.0$ &$82.0$ \\
$e^{+}e^{-} \to Z^{+}_{c1}(4050)Z^{-}_{c2}(4250)+c.c.$ & $10.867$&$-17.5\pm8.6$  &9.4  &9.1  &-&28.5&$-55.7\pm31.6$ &$30.8$ \\\hline
$e^{+}e^{-} \to Z^{+}_{c}(4050)Z^{-}_{c}(4050)$ & $10.52$&$9.4\pm15.5$  &18.1  &15.0  &1.1&23.4&$24.5\pm40.8$ &$47.7$ \\
$e^{+}e^{-} \to Z^{+}_{c}(4430)Z^{-}_{c}(4430)$ & $10.52$&$-9.7\pm8.4$  &10.5  &15.0  &-&16.9&$-25.3\pm22.3$ &$29.7$ \\
$e^{+}e^{-} \to Z^{+}_{c}(4050)Z^{-}_{c}(4430)+c.c.$ & $10.52$&$6.5\pm7.2$  &18.7  &7.5  &0.9&17.3&$33.9\pm38.0$ &$97.9$ \\
$e^{+}e^{-} \to Z^{+}_{c}(4050)Z^{-}_{c}(4050)$ & $10.58$&$7.7\pm9.3$  &23.5  &15.0  &0.7&16.5&$2.5\pm3.0$ &$7.6$ \\
$e^{+}e^{-} \to Z^{+}_{c}(4430)Z^{-}_{c}(4430)$ & $10.58$&$-60.5\pm27.8$  &22.9  &14.6  &-&12.7&$-20.1\pm9.6$ &$8.3$ \\
$e^{+}e^{-} \to Z^{+}_{c}(4050)Z^{-}_{c}(4430)+c.c.$ & $10.58$&$22.8\pm17.2$  &48.5  &7.3  &1.3&19.5&$15.1\pm11.8$ &$32.2$ \\
$e^{+}e^{-} \to Z^{+}_{c}(4050)Z^{-}_{c}(4050)$ & $10.867$&$-8.0\pm3.4$  &5.2  &14.2  &-&20.8&$-16.1\pm7.6$ &$10.8$ \\
$e^{+}e^{-} \to Z^{+}_{c}(4430)Z^{-}_{c}(4430)$ & $10.867$&$2.7\pm8.2$  &16.7  &14.0  &0.3&22.1&$5.5\pm16.7$ &$35.2$ \\
$e^{+}e^{-} \to Z^{+}_{c}(4050)Z^{-}_{c}(4430)+c.c.$ & $10.867$&$-3.7\pm5.7$  &9.1  &7.0  &-&21.1&$-15.1\pm23.4$ &$39.1$ \\
\hline
\end{tabular}
\end{threeparttable}
\end{table*}

\section{\boldmath Systematic Errors}
\label{secsys}

The following sources of systematic errors are taken into account in the
branching fraction and Born cross section measurements.

The systematic uncertainty due to charged-track reconstruction is
determined from a study of partially reconstructed $D^{\ast+}\to
D^0(\to K_S^0\pi^+\pi^-)\pi^+$ decays and is 0.35\% per
track. Based on
the measurements of the particle identification efficiencies of
lepton pairs from $\gamma\gamma\to \ell^+\ell^-$ events and pions
from a low-background sample of $D^\ast$ events, the MC simulation
yields uncertainties of $3.6\%$ for each lepton pair and $1.3\%$
for each pion. The photon reconstruction contributes 2.0\%
per photon, as determined from radiative Bhabha events.

The MC statistical errors are estimated using the yields of selected and generated events;
these are $1.0\%$ or less. Errors on the branching fractions of the intermediate states are
taken from Ref.~\cite{C38.090001}. The uncertainties of the
branching fractions for $\jpsi \to \ell^+ \ell^-$, $\chi_{c1}(1P) \to
\gamma J/\psi$, and $\psi(2S) \to \pi^{+} \pi^{-} J/\psi$ are
1.1\%, 3.6\% and 0.9\%, respectively. The trigger efficiency, evaluated from
simulation, is approximately 100\% with a negligible uncertainty.
We generate MC signal samples by assuming the $Z_c^+ \to \pi^+ \chi_{c1}(1P)$ decays are
$P$-wave and find the differences between $P$-wave and $S$-wave in the selection efficiencies are less than
1\% and thus neglected.

The difference in the 90\% C.L. upper limit on the signal yield when the mass and width of each $Z^{\pm}_{c}$
state are varied by $1\sigma$ is used as an estimate of the
systematic uncertainty associated with the mass and width
uncertainties. By changing the order of the background polynomial and the range
of the fit, the decay-mode-dependent relative difference in the 90\% C.L. upper limit on signal yields
is obtained and taken as the systematic error due to the uncertainty of the fit.

To estimate the systematic uncertainties due to the
recoil-mass requirements for $Z_c^+ \to \pi^+ + c\bar{c}$ ($cc=\jpsi,~\chi_{c1}(1P),~\psi(2S)$),
we optimize the recoil-mass regions again by changing the $Z_c^+$ widths by 1$\sigma$ and
take the differences in the 90\% C.L. upper limits on the signal yields as systematic uncertainties.

Changing the $s$ dependence of the cross sections of $e^{+}e^{-} \to
Z_c^+ + Z^{(\prime) -}_{c}$ from $1/s^2$ to $1/s^6$, the radiative
correction factors $(1+\delta)_{\rm ISR}$ become 0.651, 0.659, and
0.655 for $\sqrt{s}$ = 10.52, $10.58$, and $10.867$~GeV,
respectively. The differences are less than $1\%$ and thus neglected.

The uncertainties on the total numbers of $\Upsilon(1S)$ and
$\Upsilon(2S)$ events are 2.0\% and 2.3\%, respectively, which are
mainly due to imperfect simulations of the charged-track
multiplicity distributions from inclusive hadronic MC events.
Finally, the total luminosity is determined to 1.4\% precision using wide-angle
Bhabha scattering events.

All the systematic uncertainties are summarized in Table~\ref{sys1} for the measurements of $\Upsilon(1S,2S) \to Z^{+}_{c}Z^{(\prime) -}_{c}$ and $e^{+}e^{-} \to Z^{+}_{c}Z^{(\prime) -}_{c}$ at $\sqrt{s}$ = 10.52, 10.58, 10.867~GeV, where $Z^{+}_{c} \to \pi^{+} +c\bar c$ ($c\bar c=J/\psi, \chi_{c1}(1P), \psi(2S)$).
Assuming all the sources are independent and adding them in quadrature, the total systematic uncertainties are obtained for each mode.

\begin{sidewaystable}[htbp]
\caption{Relative systematic errors (\%) on the measurements of the branching fractions for $\Upsilon(1S,2S) \to Z^{+}_{c}Z^{(\prime) -}_{c}$
and the Born cross sections for $e^{+}e^{-} \to Z^{+}_{c}Z^{(\prime) -}_{c}$ at $\sqrt{s}$ = 10.52, 10.58, and 10.867 GeV, where
$Z^{+}_{c} \to \pi^{+} +c\bar c$ ($c\bar c=J/\psi, \chi_{c1}(1P), \psi(2S)$).}\label{sys1}
\tiny
\begin{tabular}{c | c c c c c c c c c | c}
\hline Mode &Tracking &Particle ID &Photon &MC stat. & Br & Res. Para. &Fit &Recoil Mass& $N_{\Upsilon(1S)/\Upsilon(2S)}$/Luminosity & SUM\\
\hline\hline
$\Upsilon(1S)/\Upsilon(2S) \to Z^{+}_{c}(3900)(\to \pi^{+} J/\psi)Z^{-}_{c}(3900)$ & 1.4 & 3.9 & - & 1.0 & 1.1 & 10.1/5.9 & 20.5/13.8 &9.6/6.1&2.0/2.3&25.2/16.9 \\
$\Upsilon(1S)/\Upsilon(2S) \to Z^{+}_{c}(4200)(\to \pi^{+} J/\psi)Z^{-}_{c}(4200)$ & 1.4 & 3.9 & - & 1.0 & 1.1 & 18.1/22.1 & 9.4/10.2 &18.1/19.2&2.0/2.3&27.8/31.4\\
$\Upsilon(1S)/\Upsilon(2S) \to Z^{+}_{c}(3900)(\to \pi^{+} J/\psi)Z^{-}_{c}(4200)+c.c.$ & 1.4 & 3.9 & - & 1.0 & 1.1 & 5.4/6.8 & 3.6/9.6 &18.8/10.9&2.0/2.3&20.5/16.8\\

$\Upsilon(1S)/\Upsilon(2S) \to Z^{+}_{c1}(4050)(\to \pi^{+} \chi_{c1}(1P))Z^{-}_{c1}(4050)$ & 1.4 & 3.9 & 2.0 & 1.0 & 3.8 & 15.3/6.0 & 35.2/36.8&14.0/8.3&2.0/2.3&41.3/38.7 \\
$\Upsilon(1S)/\Upsilon(2S) \to Z^{+}_{c2}(4250)(\to \pi^{+} \chi_{c1}(1P))Z^{-}_{c2}(4250)$ & 1.4 & 3.9 & 2.0 & 1.0 & 3.8 &  20.9/14.2& 17.8/14.5&20.0/26.6&2.0/2.3&34.5/34.0 \\
$\Upsilon(1S)/\Upsilon(2S) \to Z^{+}_{c1}(4050)(\to \pi^{+} \chi_{c1}(1P))Z^{-}_{c2}(4250)+c.c.$ & 1.4 & 3.9 & 2.0 & 1.0 & 3.8 &  6.4/3.1 & 7.7/10.1 &23.1/13.2&2.0/2.3&25.9/18.0 \\

$\Upsilon(1S)/\Upsilon(2S) \to Z^{+}_{c}(4050)(\to \pi^{+} \psi(2S))Z^{-}_{c}(4050)$ & 1.8 & 4.6 & - & 1.0 & 1.5 & 1.9/1.3 & 9.6/14.8&12.0/2.5&2.0/2.3&16.5/16.1 \\
$\Upsilon(1S)/\Upsilon(2S) \to Z^{+}_{c}(4430)(\to \pi^{+} \psi(2S))Z^{-}_{c}(4430)$ & 1.8 & 4.6 & - & 1.0 & 1.5 & 7.2/3.8 & 19.5/13.8&4.7/9.1&2.0/2.3&22.0/17.9\\
$\Upsilon(1S)/\Upsilon(2S) \to Z^{+}_{c}(4050)(\to \pi^{+} \psi(2S))Z^{-}_{c}(4430)+c.c.$ & 1.8 & 4.6 & - & 1.0 & 1.5 & 7.4/8.5 & 16.0/8.8&22.2/22.3&2.0/2.3&28.8/26.1 \\
\hline\hline
$e^{+}e^{-} \to Z^{+}_{c}(3900)(\to \pi^{+} J/\psi)Z^{-}_{c}(3900)$ at $10.52/10.58/10.867$~GeV &1.4 &3.9 & - &1.0 &1.1 &2.8/5.3/7.2 &5.6/9.8/12.1 &6.8/3.9/8.4&1.4 &10.3/12.7/17.0\\
$e^{+}e^{-} \to Z^{+}_{c}(4200)(\to \pi^{+} J/\psi)Z^{-}_{c}(4200)$ at $10.52/10.58/10.867$~GeV &1.4 &3.9 & - &1.0  &1.1 &17.8/13.1/11.9
&16.8/16.1/27.5 &23.4/28.3/13.4&1.4 &34.2/35.4/33.2\\
$e^{+}e^{-} \to Z^{+}_{c}(3900)(\to \pi^{+} J/\psi)Z^{-}_{c}(4200)+c.c.$ at $10.52/10.58/10.867$~GeV &1.4 &3.9 & - &1.0  &1.1 &5.8/7.6/5.0
&1.0/1.4/4.7 &21.6/18.6/14.1&1.4 &22.9/20.7/16.3\\
$e^{+}e^{-} \to Z^{+}_{c1}(4050)(\to \pi^{+} \chi_{c1}(1P))Z^{-}_{c1}(4050)$ at $10.52/10.58/10.867$~GeV &1.4 &3.9 & 2.0 & 1.0 & 3.8 &5.4/11.7/11.4
&21.0/15.7/20.3 &17.1/7.5/9.7&1.4 &28.3/21.9/26.0\\
$e^{+}e^{-} \to Z^{+}_{c2}(4250)(\to \pi^{+} \chi_{c1}(1P))Z^{-}_{c2}(4250)$ at $10.52/10.58/10.867$~GeV &1.4 &3.9 & 2.0 & 1.0 & 3.8 &15.7/23.1/15.4
&11.8/22.4/12.5 &25.6/32.0/17.3&1.4 &32.9/45.8/27.0\\
$e^{+}e^{-} \to Z^{+}_{c1}(4050)(\to \pi^{+} \chi_{c1}(1P))Z^{-}_{c2}(4250)+c.c.$ at $10.52/10.58/10.867$~GeV &1.4 &3.9 & 2.0 & 1.0 & 3.8 &9.1/7.9/10.4
&21.9/32.7/12.2 &22.1/29.1/22.8&1.4 &33.0/45.0/28.5\\
$e^{+}e^{-} \to Z^{+}_{c}(4050)(\to \pi^{+} \psi(2S))Z^{-}_{c}(4050)$ at $10.52/10.58/10.867$~GeV &1.8 &4.6 & - &1.0 &1.5 &6.8/3.2/12.3
&12.1/10.9/12.3 &18.1/10.7/10.2&1.4 &23.4/16.5/20.9\\
$e^{+}e^{-} \to Z^{+}_{c}(4430)(\to \pi^{+} \psi(2S))Z^{-}_{c}(4430)$ at $10.52/10.58/10.867$~GeV &1.8 &4.6 & - &1.0 &1.5 &7.2/5.3/6.1
&11.7/6.5/13.3 &8.4/7.9/15.6&1.4 &17.0/12.8/22.1\\
$e^{+}e^{-} \to Z^{+}_{c}(4050)(\to \pi^{+} \psi(2S))Z^{-}_{c}(4430)+c.c.$ at $10.52/10.58/10.867$~GeV &1.8 &4.6 & - &1.0 &1.5 &4.3/8.1/5.1
&7.5/9.6/9.7 &14.0/13.9/17.2&1.4 &17.3/19.5/21.1\\
\hline\hline
\end{tabular}
\end{sidewaystable}

\section{\boldmath conclusion}

\begin{figure*}[htbp]
\includegraphics[width=8.5cm,angle=-90]{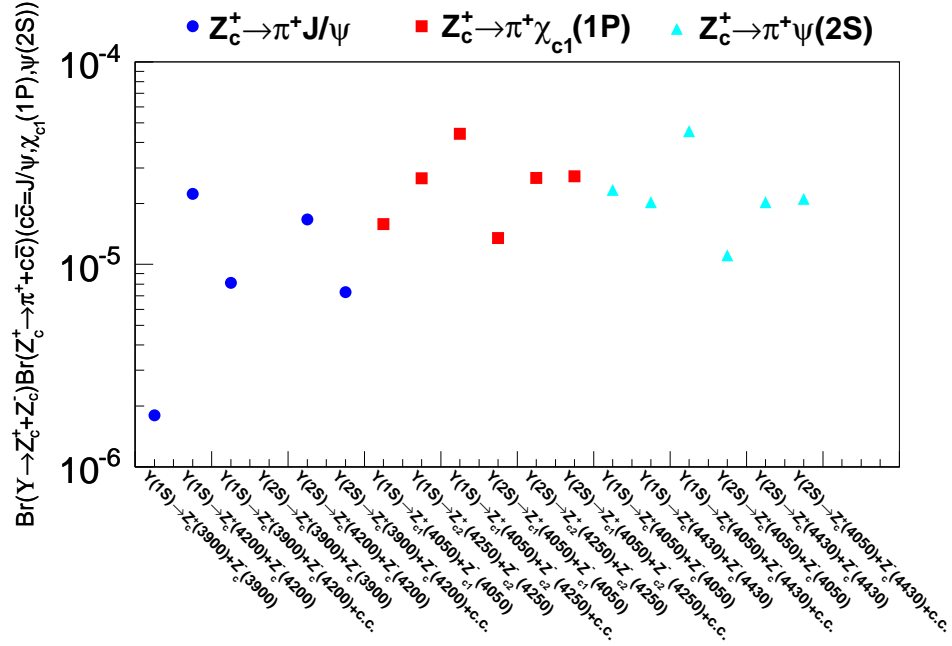}
\caption{The 90\% C.L. upper limits on $\BR(\Upsilon(1S,2S)\to
Z^{+}_{c}Z^{(\prime) -}_{c})\times\BR(Z^{+}_{c}\to \pi^{+}+c\bar c)$
($c\bar c=J/\psi,~\chi_{c1}(1P),~\psi(2S)$).
The blue circles represent the results for modes $Z^{+}_{c} \to \pi^{+}+J/\psi$, red boxes for $Z^{+}_{c} \to \pi^{+}+\chi_{c1}(1P)$ and cyan triangles for $Z^{+}_{c} \to \pi^{+}+\psi(2S)$.}\label{final1}
\end{figure*}

\begin{figure*}[htbp]
\includegraphics[width=10cm,height=15.0cm,angle=-90]{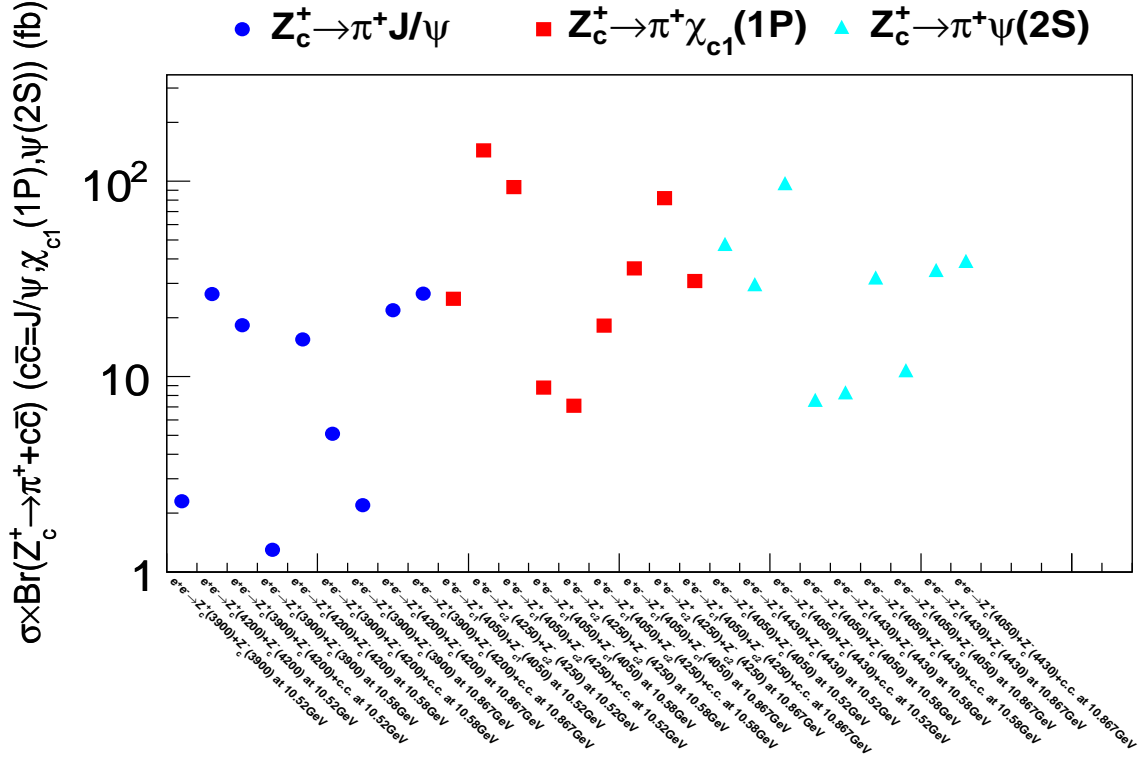}
\caption{The 90\% C.L. upper limits on $\sigma(e^+e^- \to
Z^{+}_{c}Z^{(\prime) -}_{c})\times \BR(Z^{+}_{c} \to \pi^{+}+ c\bar c)$
($c\bar c=J/\psi,~\chi_{c1}(1P),~\psi(2S)$) at $\sqrt{s}$ = 10.52, 10.58, and 10.867 GeV. The blue circles represent the results for modes $Z^{+}_{c} \to \pi^{+}+J/\psi$, red boxes for $Z^{+}_{c} \to \pi^{+}+\chi_{c1}(1P)$ and cyan triangles for $Z^{+}_{c} \to \pi^{+}+\psi(2S)$.}\label{final2}
\end{figure*}

In summary, using the large data samples of $102\times 10^6$
$\Upsilon(1S)$ events, $158\times 10^6$ $\Upsilon(2S)$ events, and
89.5 fb$^{-1}$, 702.6 fb$^{-1}$, and 121.1 fb$^{-1}$ at $\sqrt{s}$ = 10.52, 10.58, and 10.867 GeV, respectively, collected by Belle, we search
for $\Upsilon(1S,2S) \to Z^{+}_{c}Z^{(\prime) -}_{c}$ and $e^{+}e^{-}
\to Z^{+}_{c}Z^{(\prime) -}_{c}$ at $\sqrt{s}$ = 10.52, 10.58, and
10.867~GeV with $Z^{+}_{c} \to \pi^{+}+c\bar c$
($c\bar c=J/\psi,~\chi_{c1}(1P),~\psi(2S)$). No clear signals are observed
in the studied modes. We determine the 90\% C.L. upper limits on
$\BR(\Upsilon(1S,2S)\to Z^{+}_{c}Z^{(\prime) -}_{c})\times
\BR(Z^{+}_{c}\to\pi^{+}+c\bar c)$ and $\sigma(e^+e^- \to Z^{+}_{c}Z^{(\prime) -}_{c})\times \BR(Z^{+}_{c} \to \pi^+ + c\bar c)$ at $\sqrt{s}$ = 10.52, 10.58, and 10.867 GeV.  The results are displayed graphically in Figs.~\ref{final1} and ~\ref{final2}.
Due to G-parity conservation, our studied processes are electromagnetic, \textit{i.e.,} can only proceed through a virtual photon, which then transforms   into a light-quark pair ($u \bar u$) or ($d \bar d$). In this case, the dynamical suppression is much larger due to the production of two ($c \bar c$) pairs. The expected production cross section should be much lower than the double charmonium processes like $e^+e^- \to J/\psi \eta_{c}$~\cite{70.071102}. The reported upper limits are not in contradiction with the naive expectation.

\section{\boldmath ACKNOWLEDGMENTS}
We thank the KEKB group for the excellent operation of the
accelerator; the KEK cryogenics group for the efficient
operation of the solenoid; and the KEK computer group,
the National Institute of Informatics, and the
Pacific Northwest National Laboratory (PNNL) Environmental Molecular Sciences Laboratory (EMSL) computing group for valuable computing
and Science Information NETwork 5 (SINET5) network support.  We acknowledge support from
the Ministry of Education, Culture, Sports, Science, and
Technology (MEXT) of Japan, the Japan Society for the
Promotion of Science (JSPS), and the Tau-Lepton Physics
Research Center of Nagoya University;
the Australian Research Council;
Austrian Science Fund under Grant No.~P 26794-N20;
the National Natural Science Foundation of China under Contracts
No.~11435013,  
No.~11475187,  
No.~11521505,  
No.~11575017,  
No.~11675166,  
No.~11705209, and  
No.~11761141009;
Key Research Program of Frontier Sciences, Chinese Academy of Sciences (CAS), Grant No.~QYZDJ-SSW-SLH011; 
the  CAS Center for Excellence in Particle Physics (CCEPP); 
Fudan University Grant No.~JIH5913023, No.~IDH5913011/003, 
No.~JIH5913024, No.~IDH5913011/002;                        
the Ministry of Education, Youth and Sports of the Czech
Republic under Contract No.~LTT17020;
the Carl Zeiss Foundation, the Deutsche Forschungsgemeinschaft, the
Excellence Cluster Universe, and the VolkswagenStiftung;
the Department of Science and Technology of India;
the Istituto Nazionale di Fisica Nucleare of Italy;
National Research Foundation (NRF) of Korea Grants No.~2014R1A2A2A01005286, No.2015R1A2A2A01003280,
No.~2015H1A2A1033649, No.~2016R1D1A1B01010135, No.~2016K1A3A7A09005 603, No.~2016R1D1A1B02012900; Radiation Science Research Institute, Foreign Large-size Research Facility Application Supporting project and the Global Science Experimental Data Hub Center of the Korea Institute of Science and Technology Information;
the Polish Ministry of Science and Higher Education and
the National Science Center;
the Ministry of Education and Science of the Russian Federation and
the Russian Foundation for Basic Research;
the Slovenian Research Agency;
Ikerbasque, Basque Foundation for Science, Basque Government (No.~IT956-16) and
Ministry of Economy and Competitiveness (MINECO) (Juan de la Cierva), Spain;
the Swiss National Science Foundation;
the Ministry of Education and the Ministry of Science and Technology of Taiwan;
and the United States Department of Energy and the National Science Foundation.


\begin{thebibliography}{**}
\bibitem{charge} Throughout this analysis, for any given mode, the corresponding charge-conjugate mode is implied.
\bibitem{C.L.} In common high-energy physics usage, this Bayesian
interval has been reported as the ``confidence interval,''
which is a frequentist-statistics term.
\end{thebibliography}
\end{document}